\documentclass[oldversion]{aa}
\usepackage{graphicx}
\usepackage{epsfig}
\usepackage{txfonts}
\usepackage[figuresright]{rotating}
\usepackage{natbib}
\usepackage{aalongtable,lscape}

\def\cgs{erg\,cm$^{-2}$\,s$^{-1}$}
\begin{document}

   \title{The XMM-Newton Survey in the Marano Field}

   \subtitle{I. The X-ray data and optical follow-up\thanks{Based on
    observations obtained at the European Southern Observatory, Paranal, Chile
   (ESO programmes 66.B-0127(A) and 70.A-0651(A) and on observations obtained
    with XMM-Newton, an ESA science mission with instruments and contributions
    directly funded by ESA Member States and NASA.}}

   \author{M. Krumpe
          \inst{1}
          \and
          G. Lamer\inst{1} 
          \and
          A.D. Schwope\inst{1}
          \and
          S. Wagner\inst{2}
          \and
          G. Zamorani\inst{3}
          \and
          M. Mignoli\inst{3}
          \and
          R. Staubert\inst{4}
          \and
          L. Wisotzki\inst{1}
          \and
          G. Hasinger\inst{5}}

   \offprints{M. Krumpe}

   \institute{Astrophysikalisches Institut Potsdam,
              An der Sternwarte 16,
              14482 Potsdam, Germany\\
              \email{mkrumpe@aip.de}
         \and
             Landessternwarte Heidelberg-K\"onigstuhl, 69117 Heidelberg,
             Germany
         \and
             INAF - Osservatorio Astronomico di Bologna, Bologna, Italy
         \and
             Universit\"at T\"ubingen, Institut f\"ur Astronomie und
              Astrophysik, 72076 T\"ubingen, Germany
         \and
             Max-Planck-Institut f\"ur extraterrestrische Physik,
             Giessenbachstrasse, Postfach 1312, 85741 Garching, Germany}

   \date{Received 16 June 2006 / Accepted 7 December 2006} 
   \abstract{
    We report on a medium deep XMM-Newton survey of the Marano
    Field and optical follow-up observations.
    The mosaicked XMM-Newton pointings in this optical quasar survey field
    cover 0.6 deg$^2$  with a total of 120 ksec good observation time.
    We detected 328 X-ray sources in total. The turnover flux of our sample is
    $f_{\rm{X}}\sim 5 \times 10^{-15}$\,erg\ cm$^{-2}$\ s$^{-1}$ in
    the 0.2-10 keV band.
    With VLT FORS1 and FORS2 spectroscopy we classified
    96 new X-ray counterparts.

    The central 0.28 deg$^2$, where detailed optical
    follow-up observations were performed, contain 170
    X-ray sources (detection likelihood \mbox{$ML>10$}), out of which 
    48 had already been detected by ROSAT.
    In this region we recover 23 out of 29 
    optically selected quasars.
    With a total of 110 classifications in our core sample we
    reach a completeness of $\sim$65\%.
    About one third of the XMM-Newton
    sources is classified as type II AGN
    with redshifts mostly below 1.0.
    Furthermore, we detect five high redshift type II AGN
    ($2.2\le z\le2.8$).

    We show that the true redshift distribution of type II AGN 
    remains uncertain, since their lack of emission lines in a wide optical
    wavelength range hampers 
    their identification in the redshift range $1<z<2$.
    The optical and X-ray colors of the core sample indicate that 
    most of the still unidentified X-ray sources are likely to
    be type II AGN. We calculate absorbing column densities 
    and show that the ratio
    of absorbed to unabsorbed objects is significantly higher for type II AGN than for
    type I AGN.
    Nevertheless, we find a few unabsorbed type II AGN. The X-ray hardness
    ratios of some high redshift type I AGN also give an indication of
    heavy absorption. However, none of these type I objects is bright 
    enough for spectral extraction and detailed model fitting. 
    Type I and II AGN cover the same range in intrinsic X-ray luminosity,
    \mbox{($10^{43}\le L_{\rm{X}} \le 10^{46}$)}, although type II AGN have a lower 
     median intrinsic X-ray luminosity (log\,$L_{\rm{X}} \sim 44.0$)
     compared to type I AGN \mbox{(log\,$L_{\rm{X}} \sim 44.4$).}
   
    Furthermore, we classified three X-ray bright optically normal 
    galaxies (XBONGs) as counterparts. They show properties similar to
    type II AGN, probably harbouring an active nucleus.
   }

 
  \maketitle
 
\section{Introduction}
X-ray surveys are  essential to characterize the source population of
the X-ray sky and are the only means   to understand the nature of the X-ray
background radiation. Deep X-ray surveys had been carried out
during the last decades (\citealt{hasinger}, \citealt{lehmann}, \citealt{alexander}) and showed
that up to 80\% of the soft X-ray background is due to active galactic
 nuclei (AGN). XMM-Newton and CHANDRA, thanks to their high sensitivity
at hard X-rays, opened the absorbed X-ray universe for further studies
and revealed a large population of obscured, low-luminosity, low-redshift
X-ray sources (\citealt{hasinger1}).

In order to characterize the X-ray sky properly and understand the X-ray
background peak at $\sim$30 keV (\citealt{worsley}) numerous surveys with
a large variety in limiting flux and survey area  are needed
(see Fig.~1 in \citealt{brandt}). In this paper, we present new XMM-Newton
data and spectroscopic classifications of X-ray sources in the Marano Field.

The Marano Field was named by an early optical quasar survey
up to a limiting magnitude of $B_{\rm{J}}=22.0$ by
\cite{marano}. Based on different optical selection techniques
(color-color diagrams, grism plates, variability analysis) they
discovered 23 broad emission line quasars and defined an extensive
list of quasar candidates. \cite{zitelli} completed this work by
presenting a spectroscopically complete sample of quasars with
$B_{\rm{J}} \leq 22.0$ using this list of quasar candidates.
They confirmed 54 quasars including the former 23 quasars. Between
December 1992 - July 1993 ROSAT observed the central Marano Field
($\sim$0.2\,deg$^2$) for 56 ksec (\citealt{zamorani}). The data revealed
50 X-ray sources with a limiting flux of 
\mbox{$f_{\rm{X}} \geq 3.7 \times 10^{-15}\,$erg\,cm$^{-2}$\,s$^{-1}$} 
in the ROSAT band \mbox{(0.5-2.0 keV).}
Multi-color CCD and spectroscopic data identified 42 X-ray sources
(33 quasars, 2 galaxies, 3 clusters, and 4 stars). 66\% of the
optically selected quasars within the ROSAT field-of-view were detected
as ROSAT X-ray sources. \cite {gruppioni} carried out a deep radio
survey at 1.4 and 2.4 GHz and detected 68 radio sources
($f_{\rm{R}} > 0.2$~mJy). Follow-up observation provided redshifts
for 30 objects.

Our new XMM-Newton data comprise a three times larger area compared to the ROSAT
survey and are thus almost comparable in size to the optical quasar survey.
We are reaching a survey sensitivity of 
\mbox{$f_{\rm{X}} \sim 5 \times 10^{-15}\,$erg\,cm$^{-2}$\,s$^{-1}$} (turnover 
flux) over a contiguous area of 0.6\,deg$^2$. The
XMM-Newton survey of the Marano Field is thus
comparable with deep ROSAT surveys (e.g.~the UDS, \citealt{hasinger})
or with medium deep CHANDRA surveys (e.g. ChaMP, \citealt{green}).

The paper is organized as follows. In Sect.~\ref{section2} we list the
X-ray and optical data and the reduction of the data. 
Sect.~\ref{section3} describes and summarizes the 
spectroscopic classification of the X-ray sources.

In Sect.~\ref{section4} we make use of the spectroscopic classification and
redshift determination to analyze the properties 
of different object classes. In this section we concentrate 
on a 'core sample' of objects in the central part of the field, where
we reached  the highest degree of completeness in the spectroscopic 
classification.

Sect.~\ref{section6} addresses additional objects in the Marano Field that
are not X-ray detected. These objects were obtained as a control sample. 
Sect.~\ref{section5} dicusses the  results on the different
object classes. Finally, our conclusions are outlined in
Sect.~\ref{section7}.

Unless mentioned otherwise, all errors refer to a $68\%$ confidence interval.


\section{Observations and data reduction\label{section2}}

\subsection{XMM-Newton X-ray observations\label{xray}}

\begin{figure}
   \centering 
   \resizebox{\hsize}{!}{ 
   \includegraphics[bbllx=85,bblly=248,bburx=480,bbury=519,clip=]{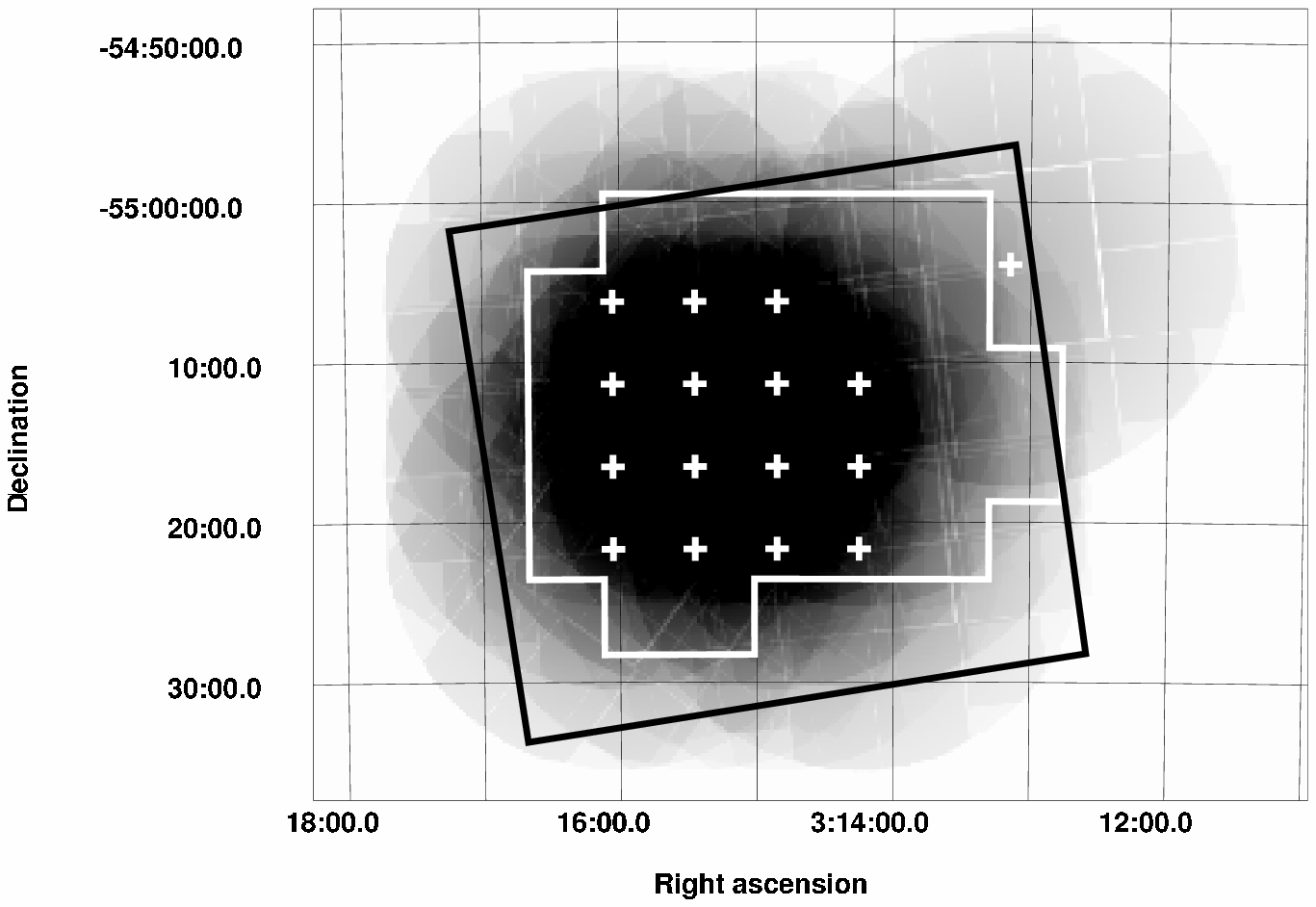}} 

      \caption{Averaged mos1 and mos2 exposure map (0.5-2.0 keV) of the   
              $4 \times 4$ pointing pattern (small white crosses). The exposure time in 
              the central region reaches 78 ksec. 
              The black rectangle shows the area covered by the 
              WFI $R$-band images. The white polygon marks the area covered 
              by the SOFI $K$-band images (see Sect.~\ref{opticaldata}).} 
         \label{exposure_map} 
\end{figure}

Since the area of the optical quasar survey in the Marano Field is larger
than the XMM-Newton field of view, the X-ray observations have been
performed as a grid of $4 \times 4$ overlapping pointings with a spacing
of $\sim$5 arcmin in right ascension and declination.
The pointing in the north-western corner of the grid was shifted in order
to cover the position of a deep far-infrared survey with the ISO satellite
(Table~\ref{obstable} and Fig.~\ref{exposure_map}).
The ISO data are not addressed in this paper. 
 
\begin{table} 
\begin{center} 
\caption{XMM-Newton observations of the Marano Field.
         Filter: T.-thin, M.-medium.}\label{obstable}  
 \begin{tabular}{ccccrc} \hline 
   XMM-orbit/ & date& RA & DEC & time&filter \\ 
         ObsID          &  {\tiny(2000)} & {\tiny (3h:min:ss)} & {\tiny (-55:min:ss)} & (ks)  &\\ 
  \hline \hline 
   106/0112940201 & 7/7 & 15:27 & 06:13 & 10.3 & T.\\ 
   107/0112940201 & 10/7 & 16:03 & 21:41 & 10.4 & T.\\ 
   107/0112940301 & 10/7 & 15:27 & 11:22 & 10.6 & T.\\ 
   107/0112940401 & 10/7 & 15:27 & 16:32 & 10.5 & T.\\ 
   107/0112940501 & 10/7 & 15:27 & 21:41 &  7.8 & T.\\ 
   129/0129321001 & 22/8 & 16:03 & 11:22 &  6.1 & T.\\ 
   129/0129320801 & 22/8 & 16:03 & 16:32 & 10.6 & M.\\ 
   129/0129320901 & 22/8 & 16:03 & 06:13 & 10.6 & T.\\ 
   130/0110970101 & 24/8 & 13:09 & 03:54 & 10.9 & T.\\ 
   130/0110970201 & 24/8 & 14:15 & 11:22 &  9.1 & T.\\ 
   130/0110970301 & 24/8 & 14:15 & 16:32 &  9.1 & T.\\ 
   130/0110970401 & 24/8 & 14:15 & 21:41 &  9.1 & T.\\ 
   131/0110970501 & 26/8 & 14:51 & 06:13 &  9.1 & T.\\ 
   131/0110970601 & 26/8 & 14:51 & 11:22 &  9.9 & T.\\ 
   133/0110970701 & 30/8 & 14:51 & 16:32 &  9.9 & T.\\ 
   133/0110970801 & 30/8 & 14:51 & 21:41 &  8.8 & T.\\\hline 
 \end{tabular} 
 
\end{center}  
\end{table}

Due to the overlapping pointings some deviations from the standard XMM-Newton  
data analysis procedures were necessary, which are described here. 
 
The photon event tables from the 16 grid pointings were merged into 
a single event table with a common sky coordinate frame using the {\tt XMM  
SAS} {\tt merge} task. In this coordinate frame we created 240 images  
(one for each pointing, energy range, and instrument). The 5 bands  
used are 0.2-0.5 keV, \mbox{0.5-2.0 keV,} 2.0-4.5 keV, 4.5-7.5 keV, 7.5-12.0 keV. 
Each image has $963 \times 963$ pixels with four arcseconds binning. 
For each of these images separate exposure maps and background  
maps were created using the {\tt XMM SAS} tasks {\tt eexpmap} and {\tt esplinemap}. 
One of the pointings (ObsID 0129320801) had accidentally been scheduled 
with a 'medium' thickness filters, while all other pointings had been  
observed with 'thin' filters. In order to correct for the somewhat  
lower throughput of the medium filters at soft energies, we multiplied  
all exposure maps of this pointing with a correction factor derived  
from the energy conversion factors for the 'thin' and 'medium' filters. 
These ECFs were taken from the SSC  document SSC-LUX-TN-0059 v.3  
(\citealt{osborne}). 
The correction is largest in the softest band (0.2-0.5 keV), where
the effective exposure in the affected pointing is reduced by 11\%. 
 
We then added the  images, exposure maps, and background maps of the  
individual pointings, resulting in 15 images for 5 energy bands and  
3 detectors. These images were used simultaneously as input for a  
detection run with {\tt eboxdetect} and {\tt emldetect}.  
The task  {\tt emldetect} applies a PSF-fit to each source found by  
{\tt eboxdetect} in order to determine the source parameters. 
Since we use merged mosaic images of the field, each source image results from
the superposition of several pointings. In each pointing the source is detected
at a different position on the detector.
Therefore, the standard configuration of {\tt emldetect}, which uses the  
off-axis angle and position-angle of each source to extract the appropriate  
PSF from calibration files, could not be used here. Instead {\tt emldetect}  
was modified to use the calibration PSF corresponding to an off-axis  
angle $\theta = 5\,\rm{arcmin}$ throughout the entire field.
This PSF is circular symmetric and is a good representation of the point  
sources in the merged images. 
For this work no extent models were fitted to the sources, therefore
the source list produced by {\tt emldetect} does not contain 
any information whether a source is extended or point-like.   
Generally, the X-ray data reduction and  
analysis was performed with {\tt XMM SAS} version 5.1 (from 18/06/2001) with  
the abovementioned exception of an adapted version of {\tt emldetect}.    
The version of  {\tt emldetect} used here suffered from an error, which  
resulted in the overestimation of detection likelihood values  
(XMM-Newton-NEWS \#29, \mbox{11-Mar-2003).} All likelihood values quoted here have  
been corrected to adhere to the relation $ML = -ln(P)$, where $ML$ is the  
detection likelihood and $P$ is the probability that the detection is caused 
by a random fluctuation of the background.  
\begin{figure}  
   \centering 
   \resizebox{\hsize}{!}{ 
   \includegraphics[bbllx=51,bblly=224,bburx=520,bbury=543,clip=]{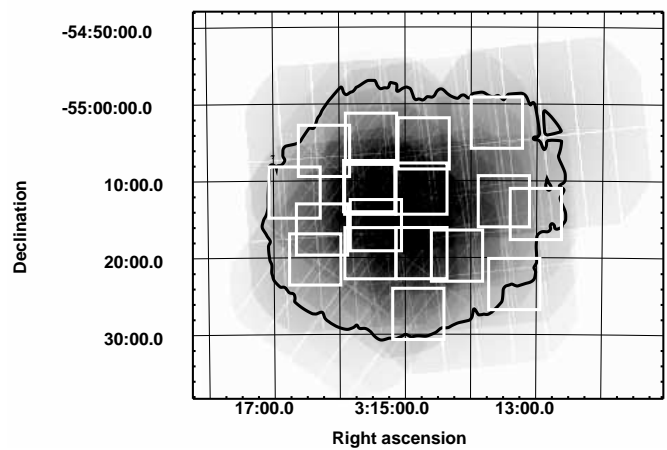}}  
      \caption{PN exposure map (0.5-2.0 keV) of the    
              $4 \times 4$ pointing pattern. The exposure time in 
              the central region reaches 35.5 ksec. The   
              area inside the black contour markes the 'core sample' 
              region where the pn-detector exposure time exceeds six ksec. 
              White rectangles show  
              the position of the spectroscopic masks for the optical 
              follow-up observations (see Sect.~\ref{spectroobs}).} 
         \label{exposure_map2}
\end{figure} 
All count rates and derived quantities are taken from the PSF fitting of {\tt 
emldetect}. 
Using the combined expoure maps, {\tt emldetect} corrects the source count rates for all 
spatial variations of telescope and detector efficiency, i.e. the count rates relate
to the optical axis of each EPIC camera.
Following \cite{cash} the 68\% confidence intervals for the 
source positions and source fluxes were calculated
using {\tt emldetect} as follows: 
Each parameter is varied until the statistic

$ C = 2  \sum\limits_{i=1}^{N}(e_i-n_i \ln e_i) $

exceeds the best fit value of $C$ by 1,
where $e_i$ is the source model at the position of pixel $i$
and  $n_i$ is the number of counts in pixel $i$.

We used energy conversion factors (ECFs) to calculate source fluxes  
in the band 0.2-10 keV (from the total EPIC count rate in the band  
0.2-12 keV) and in the band 0.5-2.0 keV  
(from the EPIC count rate in the band 0.5-2.0 keV). 
The ECFs were derived by using {\tt XSPEC} and the {\tt XMM SAS} calibration  
files to simulate power law spectra with photon index $\Gamma=1.8$ and  
the galactic column density of the field  
$N_{\rm H}=2.7 \times 10^{20}\,{\rm cm^{-2}}$. 
 
For all sources 3 hardness ratios of the form 
\mbox{$HR = (CR_2 - CR_1)/(CR_1 + CR_2)$} were calculated between the adjacent 
pn-detector energy bands 0.2-0.5 keV, 0.5-2.0 keV, 2.0-4.5 keV, 
4.5-7.5 keV.  For example $HR2$ was calculated between the bands
0.5-2.0 keV and 2.0-4.5 keV.
The different energy response of the mos- and pn-detectors  
result in different hardness ratios for the same source. 
If not noted otherwise the pn-values are used throughout the paper.
 
\begin{figure}
   \centering 
   \resizebox{\hsize}{!}{ 
   \includegraphics[bbllx=90,bblly=369,bburx=556,bbury=695]{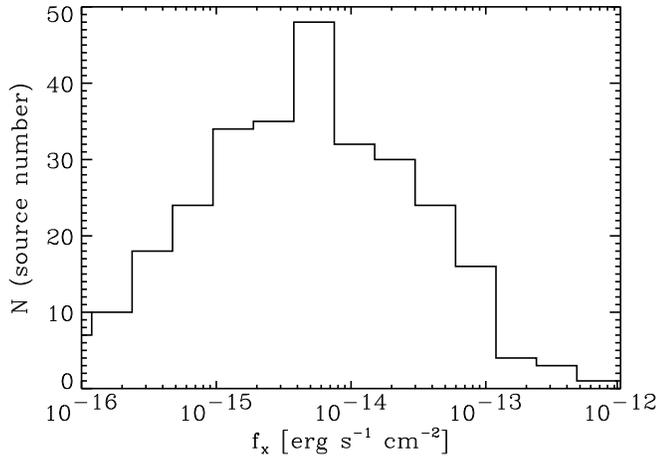}} 
      \caption{0.2-10 keV X-ray flux histogram of the 328 XMM-Newton detected
   X-ray sources in the Marano Field. Up to an X-ray flux of 
   $f_{\rm{X}}\sim 5 \times 10^{-15}$\,erg\ cm$^{-2}$\ s$^{-1}$ (turnover flux) we detect more
   and more sources. Below that X-ray flux the detection rate decreases
   dramactically.}
         \label{fig:xflux_histogram} 
\end{figure}

In total we detected 328 X-ray sources with detection likelihoods 
$ML\ge5.0$. The X-ray fluxes are in the range 
\mbox{$f_{\rm{X}}= (0.16 - 54) \times 10^{-14}$\,erg\ cm$^{-2}$\ s$^{-1}$ 
(0.2-10\,keV).} 
The X-ray flux histogram (Fig.~\ref{fig:xflux_histogram}) shows the
X-ray detection limit of our survey. Below $f_{\rm{X}}\sim 5 \times
10^{-15}$\,erg\ cm$^{-2}$\ s$^{-1}$ we are unable to detect the majority of
the X-ray sources.

In the central  
region (Fig.~\ref{exposure_map2}) we found 252 X-ray sources. The complete  
X-ray source list can be found in  
the Online Material, Appendix~A.  

Since the PSF-modelling task {\tt emldetect} was only used with 
point source models here, the detection procedure was not very
sensitive to diffuse emission from clusters of galaxies.
However, visual inspection of the X-ray images revealed no 
obviously extended sources.

We estimate that extended cluster emission down to \mbox{0.5-2.0 keV} fluxes of
$2 \times 10^{-14} {\rm\,erg\,cm^{-2}\, s^{-1}}$ should be detectable in this
survey.
The cluster log$\,N$(log$\,S)$ of \cite{rosati} 
gives a surface density of $\sim$6 clusters per square degree at this 
flux. We would therefore expect about 4 clusters in the total area
of the XMM observations and about 2 clusters in the 0.28 deg$^2$
core region of the survey.
Given the considerable cosmic variance of the cluster surface density 
the non-detection of clusters in the survey is consistent with
the log$\,N$(log$\,S)$.
In any case the  small number of expected clusters does not
suggest a significant contribution of clusters of galaxies to our source
sample. 

\begin{figure}
   \centering 
   \resizebox{\hsize}{!}{ 
   \includegraphics[bbllx=60,bblly=374,bburx=555,bbury=701]{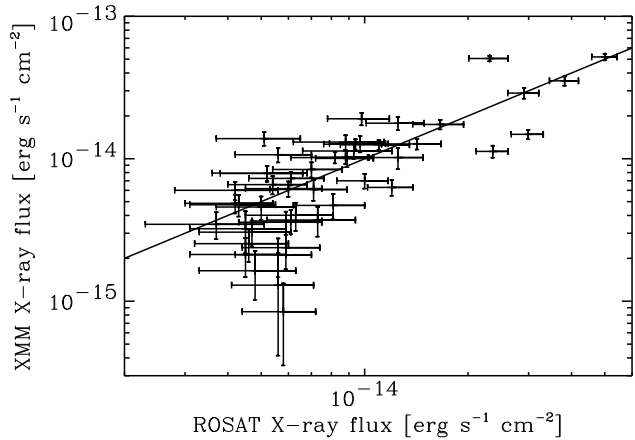}} 
      \caption{Comparison between the 0.5-2 keV X-ray fluxes measured by 
   ROSAT and XMM-Newton of all 50 ROSAT X-ray sources in the Marano Field.
   XMM-Newton fluxes are averages of the three individual EPIC camera fluxes.
   The black solid line represents equal ROSAT and XMM-Newton fluxes.} 
         \label{fig:flux_flux} 
\end{figure}

XMM-Newton redetected all 50 ROSAT X-ray sources (\citealt{zamorani}).  
However, two ROSAT X-ray sources have low detection likelihoods 
and were thus not included in the final XMM-Newton 
X-ray source list (Online Material, Appendix~A).  
Fig.~\ref{fig:flux_flux} compares ROSAT and EPIC X-ray fluxes in the 
0.5-2.0 keV band.

The X-ray fluxes of both missions are comparable down to ROSAT
fluxes of $f_{\rm{X}} \sim 6.0 \times 10^{-15}$\,erg\,cm$^{-2}$\,s$^{-1}$.  
Among the 13 objects (broad emission line objects) 
with \mbox{$f_{\rm{X}} \ge 10^{-14}$\,\cgs \ (ROSAT)} five show variability of about
a factor of two in both directions.
 
Throughout the whole paper we refer to the X-ray sources by 
mentioning the X-ray source number without any letter as suffix (e.g. 452). 
These numbers refer to the sequence numbers in the original {\tt emldetect} 
source list and are non-contiguous due to the removal of sources below  
$ML = 5.0$ after correction of the detection likelihoods.

\subsection{The optical data\label{opticaldata}}

\subsubsection{WFI $R$-band observations} 
 
\begin{figure}
   \centering 
   \resizebox{\hsize}{!}{ 
   \includegraphics[bbllx=72,bblly=369,bburx=545,bbury=695]{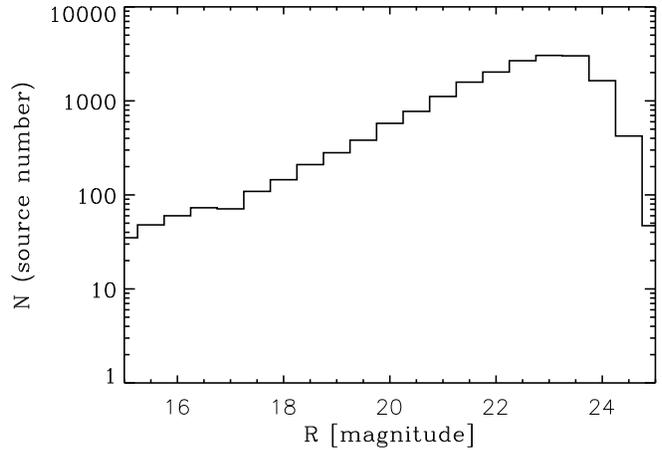}} 
      \caption{Magnitude histogram of the sources detected in the WFI $R$-band
   image. Completeness in detecting objects is
   lost for $R>23$.}
         \label{fig:magR_histogram} 
\end{figure}

The central region of the Marano Field has been observed  
in the $UBVRI$ bands with the ESO  
Wide-Field-Imager (WFI). 
Here we make use of the deep (7000 sec) $R$-band image (\citealt{mignoli}) with a 
limiting magnitude $R \simeq 22.5$ (defined as turnover magnitude minus 0.5 mag, see Fig.~\ref{fig:magR_histogram}).
The WFI-image covers an area of \mbox{35 $\times$ 32 arcmin}  
well aligned with the region of the deepest XMM-Newton coverage 
(Fig.~\ref{exposure_map}).  

\subsubsection{SOFI $K$-band observations} 
\begin{figure}
   \centering 
   \resizebox{\hsize}{!}{ 
   \includegraphics[bbllx=72,bblly=369,bburx=545,bbury=695]{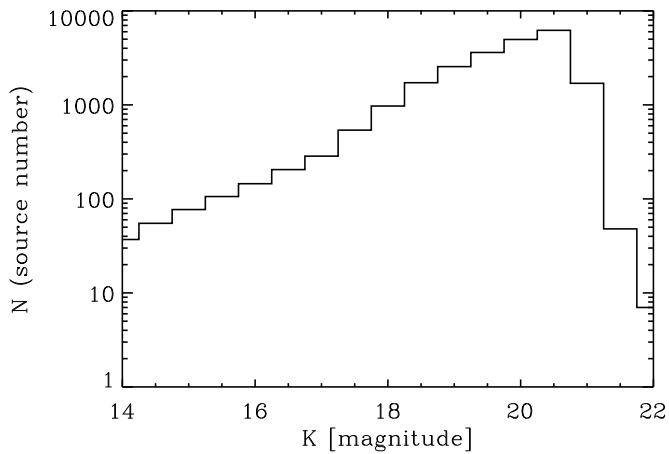}} 
      \caption{Histogram of the $K$-band magnitudes in the NTT-SOFI images.
   Completeness in detecting objects is
   lost for $K>20.5$.}
         \label{fig:magK_histogram} 
\end{figure}

The $K$-band data ($2.0-2.3\,\mu$m) were obtained at the ESO New  
Technology Telescope (NTT) with the SOFI instrument.  
With an area of 754 arcmin$^2$ the $K$-band mosaic  
covers a slightly smaller area than the WFI $R$-band observation and 
is well aligned with the deepest XMM-Newton exposure
(Fig.~\ref{exposure_map}). 
The SOFI observations consist of a mosaic of 33  
jittered pointings, each covering $5 \times 5$ arcmin 
with an exposure time of 29 minutes each.   
The limiting magnitude is $K\simeq20$ (turnover magnitude minus 0.5 mag, see Fig. \ref{fig:magK_histogram}). 
 
\subsubsection{VLT $R$-band pre-images} 
$R$-band pre-images for the spectroscopic follow-up observation 
were obtained with the Focal Reducer and Spectrograph FORS1 and FORS2  
at the ESO Very Large Telescope (VLT). The first  
run (FORS1) contained six and the second run  
(FORS2) 12 optical images. The summed pre-imaging area  
is $\sim$530 arcmin$^2$ in the central region 
of the Marano Field (Fig.~\ref{exposure_map2}). 
Since the field was not contiguously covered by FORS spectroscopy, also the
pre-images cover only $\sim$50\% of the central area of the X-ray survey.

\subsection{Target selection for optical spectroscopy\label{target}} 
 
Our target selection for the optical spectroscopy 
was primarily based on FORS1/FORS2 $R$-band pre-images.  
WFI $R$-band and SOFI $K$-band data were used in addition  
to support the target selection on the pre-images 
in case of faint counterparts. 
The chosen fields represent a compromise in terms of 
maximum survey area and source density because of 
limited telescope time. 
First priority for spectroscopic follow up was given to candidates with a 
likelihood of existence $ML > 10$ within a 2$\sigma_{\rm{X}}$ error radius.
The total position error is \mbox{$\sigma_{\rm{X}} = \sqrt {\sigma_{\rm{stat.}}^2+\sigma_{\rm{syst.}}^2}$.}
The statistical errors $\sigma_{\rm{stat.}}$ of the positions were calculated 
as described in Sect.~\ref{xray}.
The systematic errors are caused by uncertainties in the spacecraft attitude,
errors in the linearisation of the detector coordinates, and undersampling
of the PSF (in particular for the PN camera).
At this stage, we allowed a systematic position error of 2 arcsec, see 
Sect.~\ref{falsematch} for a more accurate estimation of the systematic
position errors.

The multi-object spectroscopy masks were designed with 
the  software {\tt fims 2} provided by ESO. For all spectroscopic  
targets we used straight slitlets with a  nominal length of ten 
arcseconds and width of one arcsecond.  
For extended sources or closely spaced candidate counterparts  
a different slit length, \mbox{6 -- 14} arcseconds, was used. 
Still available mask positions were filled with 
candidate objects not fulfilling the main selection criteria and with  
additional random non X-ray emitting galaxies.

\subsection {Spectroscopic observations\label{spectroobs}} 
The FORS1 multi object spectroscopy (MOS) run  
with six masks was obtained on November, 20th in 2000 by 
using grism \emph{GRIS\_150I+17}. No order separation filter was used. 
The exposure time for every mask was 2400 seconds.  
The seeing was 0.7$\pm$0.1 arcsec. 
 
The second run was performed with FORS2 in spectroscopic 
mask mode (MXU) from November, 27th-30th in 2002.  
We aimed for an exposure time of 3 $\times$ 1800 seconds per mask,   
but weather and time constraints required some deviations from  
this general scheme (Table~\ref{tab:fors2}). Grism  
\emph{GRIS\_150I+27} was used for all masks. 
\begin{table}
\caption[]{Details of FORS2 spectroscopic mask mode observation 
           (see Fig.~\ref{exposure_map2}\label{tab:fors2})} 
\begin{center} 
\begin{tabular}{lccc}\hline 
 mask & observation  & filter & seeing \\ 
      & (each 1800s) &        & (in arcsec)\\ 
\hline
   mask1  &    3 &  none   & 0.63\\ 
   mask2  &    4 &  none   & 1.36\\ 
   mask3  &    3 &  none   & 1.05\\ 
   mask4  &    3 &  none   & 0.78\\ 
   mask5  &    3 &  none   & 0.93\\ 
   mask6  &    3 &  GC375  & 0.75\\ 
   mask7  &    3 &  none   & 0.80\\ 
   mask8  &    3 &  none   & 0.81\\ 
   mask9  &    2 &  none   & 0.66\\ 
  mask10  &    3 &  none   & 0.76\\ 
  mask11  &    1 &  none   & 0.96\\ 
  mask12  &    2 &  none   & 0.64\\\hline 
\end{tabular} 
\end{center} 
\end{table} 
In order to prevent second order contamination, the first mask (mask6)  
in the observing sequence was observed through filter GC375 which limited  
the wavelength range from 3850 to 7500 \AA.  
A comparison with the second mask without order separating filter  
showed that the contamination effect is negligible. Hence, 
for all following masks the filter was removed from the light path resulting 
in a wavelength range for central targets of the final spectra from  
\mbox{3500 to 10000 \AA}. 
 
\subsection{Spectroscopic data reduction} 
 
The reduction of the data was accomplished by  
a semiautomatic pipeline coded in {\tt MIDAS}. It was specially  
designed to reduce FORS2-MXU data 
with as little interaction by the user as possible.  
After modifications this code was also used to reduce 
the FORS1-MOS data.\\ 
The bias correction was done in the standard manner with careful 
attention to possible time dependence on the bias level and dark 
current. An ordinary flat field correction was used to rectify the 
pixel-to-pixel variation.  
  
\subsubsection{Wavelength calibration} 
We established a 2-d wavelength calibration in the pipeline. This  
allowed to correct the distortion perpendicular to  
the dispersion direction and improved significantly the  
signal-to-noise ratio of the extracted spectra.  
The calibrated wavelength range is
3500--10000 $\AA$. However, many spectra were located close to  
the edge of the CCD. Hence, there are substantial variations in the  
wavelength range of the spectra. The average uncertainty of the  
wavelength calibration is 0.2 \AA\, but reaches a maximum of 
$\Delta \lambda \pm 1.5\,\AA$ 
at long and short wavelength ends. Further details on the 
\begin{table} 
\caption[]{Setups for the optical spectroscopy}\label{tab:wave} 
\begin{center} 
\begin{tabular}{rcc}\hline
 & FORS1 & FORS2 \\\hline 
grism & GRIS\_150I+17 & GRIS\_150I+27\\ 
dispersion [$\AA /$mm]& 230&225 \\ 
pixel size [$\mu$m]& 24 x 24& 15 x 15\\ 
step width [$\AA/$Pixel]& 5 & 3 \\\hline 
\end{tabular} 
\end{center} 
\end{table} 
optical setups can be found in Table~\ref{tab:wave}.  
We estimate the spectral resolution finally achieved by  
measuring the width of the arc lines in the lamp spectra to 
$\Delta \lambda \sim 21 \AA$ (FWHM). 
 
\subsubsection{Object and sky definition} 
The aquisition slit-through images were used to roughly define the  
position and width of every single slit in an image. 
For an optimal definition of the object and sky region in the spectrum 
the pipeline then displays intensity profiles in graphic windows and 
an image of the corresponding spectrum. After a careful inspection of 
these pipeline outputs, the object and the sky region were defined manually 
for every single spectrum. Consequently, bad pixel/lines/colums in the  
raw data can be excluded for the extraction and scientific  
misinterpretation of artifacts in the final spectra can be avoided.     
Whenever possible, two sky regions, on both sides of the target spectrum, were
defined.
 
\subsubsection{Extraction of spectra} 
 
The extraction is based on an optimal extraction algorithm  
(\citealt{horne}) including a cosmic ray rejection. Extensive tests 
determined the optimal extraction parameters.  
Standard flux calibration was applied. Since FORS2-mask6 was  
observed with an additional filter, a different standard star  
was used. All spectra were corrected for atmospheric absorption using the 
standard ESO extinction correction function scaled to the given airmass. 
 
The $n$ individual spectra of a given counterpart object were combined to form 
one single final spectrum. The individual spectra were firstly normalized to 
the same mean intensity. The normalization factor  
was determined from an analysis of all spectra in a given mask. 
The final spectrum is the uncertainty-weighted mean spectrum. 
 
\section {Spectroscopic classification of the X-ray counterparts 
  \label{characterization}\label{section3}}  
The spectroscopic classification of the individual X-ray sources 
was primarily based on the FORS1 and FORS2 spectra. In 
addition, we used optical and X-ray images for a  
reliable identification (see Online Material, Appendix~C). 
Broad emission line objects were immediately accepted as  
X-ray counterparts. Narrow emission line galaxies and normal galaxies 
were accepted as X-ray counterparts, if no other optical candidate was found
within the position error range. 
Stars were regarded as likely counterparts if the X-ray
colors indicated a soft X-ray spectrum compatible  
with coronal emission of $\leq 1$ keV. 
Every X-ray identification was confirmed by at least two individuals. 

\begin{figure}
   \centering 
   \resizebox{\hsize}{!}{ 
   \includegraphics[width=12cm,angle=270]{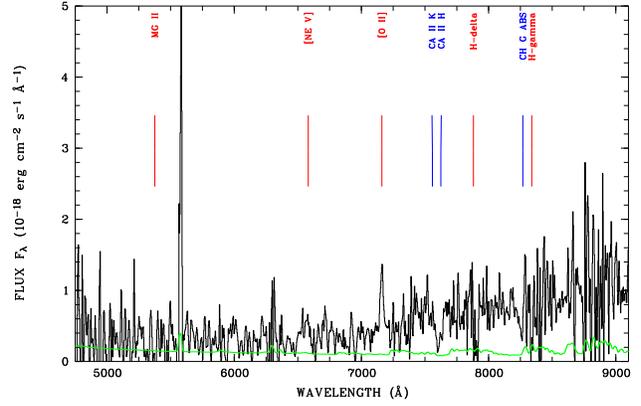}} 
      \caption{Optical atmospheric absorption corrected, wavelength and 
               flux calibrated spectrum for X-ray source 480.
               The black spectrum marks the object's spectrum. The green line
               (see the color online version of the manuscript) shows the 
                error spectrum. Possible emission and absorption
               features in the spectrum are labeled. The spectral feature at 
               5580 \AA\ is spurious due to incomplete substraction of a night
               sky line.}
         \label{fig:spectrum} 
\end{figure}

For the large majority of our spectra the signal
to noise ratio (SNR) is sufficient to give reliable classifications and redshifts.
Like other X-ray identification surveys, we encounter the problem of
difficult classification of the optical spectra below a certain SNR.
For spectra with a SNR=3-5, the identification of narrow emission 
lines was still possible. However, at this SNR faint broad emission 
lines and normal galaxy spectra are very 
difficult to identify. 
Fig.~\ref{fig:spectrum} shows a spectrum of a narrow emission line
galaxy with a SNR=2.4 (continuum near the \ion{O}{ii} emission line), which is 
close to the identification limit. Only one significant narrow emission 
line at $\sim$7100 \AA\ is found. The spectral shape makes it reasonable 
to identify this line as \ion{O}{ii} emission line. 
Even though this classification is likely, an unambiguous redshift
determination and classification of the type cannot be given. 
Spectra with a SNR less then $\sim$2.5 
are not identifiable.

The reliability of the redshift determination and classification of the 
optical spectra is given by flags in column \emph{(9) flags} of Table~\ref{opt_tab}. 

The complete list of X-ray classification is given in Table~\ref{opt_tab}.
The columns are described as follows:\\

\emph{(1) No}\\ 
Classification of a counterpart object consists of the sequence number of 
the X-ray source list and a suffix (A, B) in order to discriminate 
between different candidates.\\  
 
\emph{(2) RA [hh:min:ss] and (3) DEC [deg:min:ss]} \\ 
Right ascension and declination of the optical candidate counterpart 
(J2000).\\   
 
\emph{(4) dist$_{OX}$ [arcsec]} \\ 
Spatial offset between the X-ray and optical positions.\\
 
\emph{(5) K} \\ 
SOFI $K$-band magnitude of the spectroscopically classified candidate, whenever available.\\ 
 
\emph{(6) R}\\ 
WFI $R$-band magnitude of the spectroscopically classified candidate, whenever possible.\\ 
 
\emph{(7) class}\\ 
Spectroscopic classification of the identified object. S -- star, G -- normal galaxy 
(no emission lines), N -- narrow emission line galaxy with unresolved emission 
lines (at 6000 $\AA$ our spectral resolution of 21 $\AA$ corresponds to 1050 km/s), B -- broad emission line object  
(all measured line widths have $v_{\rm{FWHM}}>2500$\,km/s), ? -- undefined object.\\

\emph{(8) $z$}\\ 
Spectroscopic redshift of the identified object. The  
redshift is taken from the literature for objects with 
'1\,-\,-' and '0\,-\,-' in column '\emph{(9) flags}'. 
Column '\emph{(15) remarks}'
states the source of redshift determination and classification.\\

\emph{(9) $flags$}\\ 
X-ray identification flag, redshift flag, and classification reliability flag.
The first number (0,1) marks whether a spectroscopically classified object was accepted 
as X-ray counterpart. Objects which we consider to be the correct 
identification of the X-ray source are flagged by '1', while '0' flags objects
not considered as the X-ray source.\\
The second (middle) flag states the redshift reliablility. A redshift flag 
'1' means a reliable, well established redshift determined by several spectral
features. '0' marks objects where the redshift determination relies on 
a single but reasonable spectral feature.\\
The third (last) flag characterizes the classification reliability. A flag
'1' marks that the object type as given in '\emph{(7) class}' is 
well established and reliable. Flag '0' indicates an uncertain 
classification of
the object type. Either high SNR spectral features of the object do not 
allow a proper classification or the optical spectra do not allow 
to give a reliable classification of the object type because of a low SNR 
and/or insufficient wavelength coverage of the optical spectra.
The latter is illustrated in Fig.~\ref{fig:spectrum}. The narrow \ion{O}{ii}
line indicates a narrow emission line galaxy. However, the SNR of the
spectrum does not allow to judge the existence of a 
broad \ion{Mg}{ii} line and, hence, a classification as type I AGN.
Since \ion{C}{iv} and \ion{C}{iii} are outside the spectrum's 
wavelength range, \ion{Mg}{ii} is the only possible broad line feature.
The most likely classification of this object is a narrow emission line 
galaxy. However, the given arguments show that a classification as type I AGN 
cannot be excluded. Therefore, the classification flag for this object is '0'.\\
Objects with '1\,-\,-' and '0\,-\,-' have only an X-ray identification flag
since their redshift and classification relies on follow-up surveys 
previously done in the Marano Field (see column '\emph{(15) rem.}').\\

\emph{(10) log\,(L$_{X_{OBS}}$ [erg~s$^{-1}$])}\\ 
Observed rest-frame X-ray luminosity (logarithmic units) in the
0.2-10 keV energy band calculated by using Eq.~\ref{xlum}. 
The $k$-correction vanishes since we assume  
an energy index $\alpha = -1$ with  
\mbox{$F_{\rm{\nu}} \sim \nu^ {\,\alpha} \sim \nu^{\,1-\Gamma}$} ($\Gamma$ - 
photon index) based on \cite{alexander} and \cite{mainieri}. The luminosity  
distance $d_{\rm{L}}$ was computed by the analytical fit for flat  
comologies with $\Omega_{\rm{m}} =0.3$, $\Omega_{\Lambda} =0.7$,  
$H_{\rm{0}}=70$\,km\,s$^{-1}$\,Mpc$^{-1}$ following \cite{szokoly}. 
\begin{eqnarray} 
\label{xlum} 
L_{\rm{X}} = \frac{4\,\pi\,d_{\rm{L}}^{2}}{(z+1)^{\,\alpha+1}}f_{\rm{X}} 
\end{eqnarray}\\ 
 
\emph{(11) $M_B$}\\ 
Absolute magnitudes $M_{\rm{B}}$ (in the Johnson system) were estimated only
for type I AGN using the relation
\begin{eqnarray}
\label{MB}
M_{\rm{B}} = R + 5 - 5 \log (d_{\rm{L}}/\mathrm{pc}) + K(z)
\end{eqnarray}
where $d_{\rm{L}}$ is the luminosity distance 
and $K(z)$ is the customary $k$-correction term. 
\begin{figure}           
   \centering
   \resizebox{\hsize}{!}{
   \includegraphics[width=12cm,angle=270]{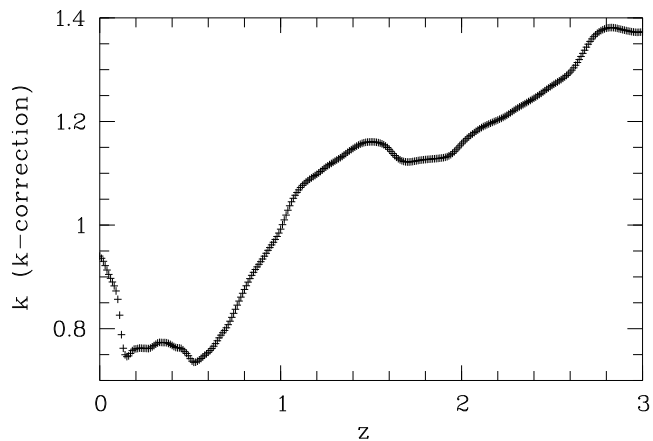}}
      \caption{Adopted mean k correction from observed R-band to rest-frame
   $M_{\rm{B}}$, for broad emission line AGN.
              }
         \label{k-corr1}
\end{figure} 
In our case, this term
includes the transition from observed-frame $R$-band to rest-frame $B$-band,
assuming a mean spectral energy distribution for all sources,
and also the $(1+z)$ bandwidth stretching factor.
For the type~I AGN we computed $K(z)$ from the composite SDSS quasar 
spectrum (\citealt{vandenberk}). The resulting graph is shown in 
Fig.~\ref{k-corr1}.\\ 
 
\begin{figure}
   \centering
   \resizebox{\hsize}{!}{
   \includegraphics[width=12cm,angle=270]{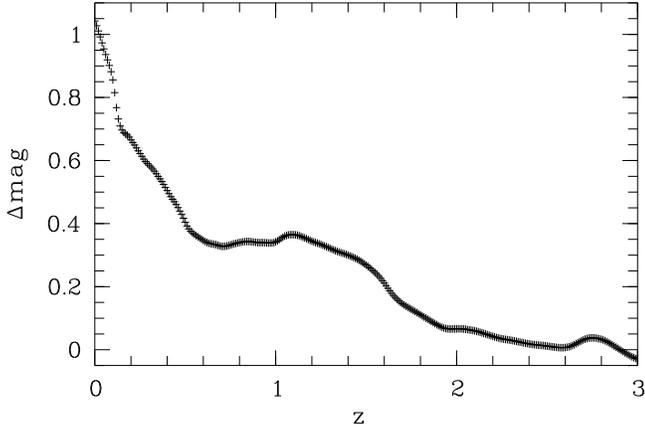}}
      \caption{Adopted conversion term to estimate AB magnitudes 
               at fixed rest-frame wavelength $\lambda = 2500$\,\AA\ from
               observed R-band magnitudes.
              }
         \label{k-corr2}
\end{figure}

\emph{(12) $\alpha_{OX}$}\\  
The broad band spectral index $\alpha_{\mathrm{OX}}$ roughly characterizes 
the UV--X-ray spectral energy distribution by connecting the
rest-frame points at 2500~\AA\ and 1 keV with a simple power law, 
$F_{\nu} \propto \nu^ {-\alpha_\mathrm{OX}}$.
For each broad emission line AGN we estimated its flux at a fixed 
rest-frame wavelength of $\lambda = 2500$~\AA, applying the relation
\begin{eqnarray}
\label{flux2500-1}
m_\mathrm{AB}(2500\,\mathrm{\AA}) = R + \Delta m(z)
\end{eqnarray}
where $R$ is the quoted $R$-band magnitude, $m_\mathrm{AB}(2500\,\mathrm{\AA})$ is the predicted
magnitude at 2500~\AA\ -- expressed in the AB system for easy conversion into
monochromatic fluxes --, and $\Delta m(z)$ is a redshift-dependent term
(similar, but not identical to the $k$-correction) that also accounts for the
zeropoint transformation from the Vega to the AB system.
Our adopted $\Delta m(z)$ relation, again computed from the SDSS composite 
quasar spectrum of \citet{vandenberk}, is shown in Fig.~\ref{k-corr2}.
Notice that at the typical redshifts of $z \sim 1.5$ of our broad line AGN,
the observed $R$-band approximately traces a rest-frame wavelength
of $\lambda_\mathrm{rest} \sim 2600$~\AA, implying that the spectral energy distribution corrections
are small. The resulting AB magnitudes are then converted into fluxes
following the definition of the AB system (\citealt{oke}):
\begin{eqnarray}
\label{flux2500-2}
m_\mathrm{AB}(2500\,\mathrm{\AA}) = -2.5 \log(F_\nu(2500\,\mathrm{\AA}))-48.60
\end{eqnarray}
where $F_\nu(2500\,\mathrm{\AA})$ is given in erg s$^{-1}$ cm$^{-2}$ Hz$^{-1}$.

The X-ray flux at 1~keV is computed by:
\begin{eqnarray}
\label{flux1kev}
f(0.2-10\,\mathrm{keV}) = F_\nu(1\,\mathrm{keV})
\int\limits_{0.2\,\mathrm{keV}}^{10\,\mathrm{keV}}
E^{\alpha} \mathrm{d}E \,\,\,\,\,\,\,\mathrm{with}\,\, \alpha = -1\,.
\end{eqnarray}
 Hence, the broad band spectral index is obtained as
\begin{eqnarray}
\label{spindex}
\alpha_{ox} = 
\frac{\log(F_\nu(2500\,\mathrm{\AA}))-\log(F_\nu(1\,\mathrm{keV}))}{\log(\nu(1\,\mathrm{keV}))-\log(\nu(2500\,\mathrm{\AA}))}.
\end{eqnarray}
\\

\emph{(13) N$_H$ [cm$^{-2}$]} \\ 
X-ray absorbing hydrogen column density in units of $10^{22}~\mathrm{cm^{-2}}$
(see Sect.~\ref{nh}).\\ 
 
\emph{(14) log\,($L_{X_{INT}}$ [erg~s$^{-1}$] )} \\ 
Intrinsic rest-frame X-ray luminosity (logarithmic units) in the  
0.2-10 keV energy band after X-ray flux correction for the absorbing  
hydrogen column density. Calculation uses Eq.~\ref{xlum}.\\ 
 
\emph{(15) rem.}\\ 
Remarks for individual objects: 1 -- optically selected and spectroscopically classified  
quasar by \cite{marano}, \mbox{2 -- optically} selected and spectroscopically classified
quasar by \cite{zitelli}, 3 -- ROSAT X-ray source with spectroscopic classification  
and redshift determination by \cite{zamorani}, 4 -- ROSAT X-ray source with 
no or wrong identification by \cite{zamorani}, 5 -- unclassified radio objects 
within 5 arcsec, \cite{gruppioni},  
\mbox{6 -- spectroscopic} classification and redshift taken from    
\cite{teplitz}, \mbox{7 -- radio} source, spectroscopic classification and redshift  
taken from \cite{gruppioni}, 
C -- individual comment to an object (see \ref{comments_objects}).\\

\subsection{Comments to individual objects\label{comments_objects}}

{\hspace{0.55cm}{\bf 9A}: New redshift determined for this source, formerly known as \cite{zamorani} 
\mbox{X043-12}. 
 
{\bf 15A}: Object 15A is not regarded as the optical counterpart, since the 
           identification of the line features is uncertain and the positional 
           offset is rather large. A very faint object at the detection 
           limit lies in the X-ray error circle. A lower limit of  
           $f_{\rm{X}}/f_{\rm{opt}}\sim 65$ for this source 
           was estimated by assuming a limiting magnitude of $R=24$.
 
{\bf 17A}: Source 17 is a likely X-ray blend with large contrast between the 
           two individual sources. Object 17A is a unique identification of 
           the brighter X-ray source, whose X-ray flux is likely to be 
           overestimated due to blending. 
 
{\bf 20A \& 20B}: \cite{zamorani} identified source 20 with the M-star 20B.  
Some contribution of the M-star's X-ray flux cannot be excluded, but the 
identification with the type II AGN (object 20A) seems more likely, since the X-ray colors 
indicate a relatively hard spectrum. 
 
{\bf 22A \& 22B}: X-ray source 22 is a probable blend. Objects 22A and 22B 
                  fall on top of the two suspected point sources. We regard 
                  both optical objects as likely counterparts of the X-ray 
                  blend and distribute the measured X-ray flux evenly between 
                  the two sources. 
 
{\bf 26A}: New redshift determined for this source, formerly known as \cite{zamorani} 
\mbox{X301-29}. 
 
{\bf 32A}: N{\sc V} strongest emission line in the spectrum (brighter than 
           Ly-$\alpha$). Narrow emission lines are present and the object is
           classified as 'N' (narrow emission line galaxy). However, all
           narrow emission lines have underlying broad components. 
           Therefore, the classification flag is set '0'.

{\bf 35A \& 35B}: Two galaxies at equal redshift, the brighter galaxy  
                  35A is regarded as the X-ray counterpart. A nature of the X-ray 
                  source as galaxy cluster cannot be completely excluded but an 
                  X-ray extent is not obvious. For 35A a new redshift is 
                  determined (update to \cite{zamorani} X022-48). 
 
{\bf 42A}: New redshift determined for this source, formerly known as  
           \cite{zamorani} \mbox{X031-24}. 
 
{\bf 46A}: X-ray blend with major contribution from the southwestern 
           component with type I AGN 46A as counterpart.  
            
{\bf 47A \& 47B}: Two narrow emission line galaxies at equal redshift. A 
                  nature of the X-ray source as galaxy 
                  cluster cannot be completely excluded, but an X-ray extent is not obvious.  
                  47A is assumed as optical counterpart. 

{\bf 63A}:  Unresolved narrow emission lines are present and the object is
           classified as 'N' (narrow emission line galaxy). However, 
           the \ion{C}{iv} emission line has an underlying broad component.
           Therefore, the classification flag is set '0'.
 
{\bf 66A}: The identification is not unique, since the optical 
           image reveals a possible second   
           object inside the 1$\sigma$-X-ray position error 
           circle. 

{\bf 133A}: Narrow emission lines are present. The object is
           classified as 'N' (narrow emission line galaxy). However,
           Ly-$\alpha$ shows a resolved, broad base. Therefore, 
           the classification flag is set '0'.

{\bf 151A}: Probably a spurious detection of the X-ray source. The source is 
kept in the source list for the formal reason of having an $ML = 5.2$ but is 
not considered further. 

{\bf 191A}: Optical spectrum in \cite{zitelli} indicates a type I AGN 
            with typical broad emission lines. However, this object shows
            the lowest X-ray to optical flux ratio of all AGN in our sample. 
            $f_{\rm{X}}/f_{\rm{opt}}=0.04$ suggests an X-ray faint AGN. 
            Since this object was not detected by ROSAT, it is unlikely
            that the extreme $f_{\rm{X}}/f_{\rm{opt}}$-ratio is due to a
            temporary low X-ray state of the object. Furthermore, it is one of
            the type I objects with intrinsic absorption
            ($N_{\rm{H}}=(5.75_{-2.81}^{+3.83})\times 10^{22}\,{\rm{cm}}^{-2}$).  
 
{\bf 217A \& 217B}: X-ray blend, the identification of 217A with the 
                    southeastern component seems unambiguous, the 
                    identification of 217B with the northwestern component is 
                    not unique, since a similarly bright, close-by, but still 
                    unidentified object is present at the same distance from 
                    the X-ray source. 
 
{\bf 224A \& 224B}: Two galaxies at equal redshift, no obvious X-ray extent.

{\bf 253A \& 253B}: Two objects at similar redshift with $2\,\mathrm{mag}$
                    difference in the
                    optical. The brighter object 253A is regarded as 
                    the identification. 

{\bf 280A}:   Object is classified as 'B' since \ion{C}{iii} is 
              well resolved with $\Delta \lambda \sim 72 \AA$ (FWHM).
              Ly-$\alpha$ and \ion{C}{iv} are narrow emision lines, but 
              show strong absorption with broad underlying components.
              The classification flag is set '0'. Possible broad absorption line quasar.

{\bf 361A}:    Broad absorption line quasar.
 
{\bf 382A \& 496A}: Physical quasar pair, separated by $17\arcsec$, at $z = 
1.904$, d$_{\rm{A}}$ = 143 kpc. The spectra are different, i.e.~the two objects are 
not lensed images of the same source. 
 
{\bf 437A}:   Spectrum, optical and X-ray image,  
              and relative soft hardness ratios point undoubtedly 
              to an M-star as X-ray source. However, a flux ratio $f_{\rm{X}}/f_{\rm{opt}}=1.7$ 
              is unusually high for an M-star as the X-ray identification.
 
{\bf 512A \& 512B}: \cite{zamorani} identify the NELG 512B with the X-ray 
                    source. The $2\,\mathrm{mag}$ fainter NELG 512A lies somewhat closer 
                    to the X-ray position. While both galaxies may contribute 
                    to the observed X-ray flux, we assume object 512A as the 
                    counterpart in the following. 
 
{\bf 582A}:   Spectrum, optical and X-ray image,  
              and relative soft hardness ratios point undoubtedly 
              to an M-star as X-ray source. However, a flux ratio $f_{\rm{X}}/f_{\rm{opt}}=1.5$ 
              is unusually high for an  M-star as the X-ray identification.
 
{\bf 607A \& 653}: Detected as a single X-ray source by ROSAT \cite{zamorani}  
                  X404-23, 
                  X-ray source 653 is brighter and closer to X404-23  
                  and, therefore, treated as the detected ROSAT X-ray source.
                  The broad spectral feature at 7100$\,\AA$ in the optical
                  spectrum of 607A is spurious due to the zeroth order light of
                  the neighbouring slit. 
 
{\bf 615A \& 615B}: Both objects, the broad emission line object 615A and the
narrow emission line 
object 615B, are possible counterparts to the X-ray source. We regard the 
fainter, but positionally better matching object 615A as the counterpart. 
                     
{\bf 632A}: Spectrum suggests a BL-Lac object, but the object is  
            not a radio source, classification unclear.

\subsection{Discussion of spurious matches \label{falsematch}}

Since type I AGN are a well-established class of X-ray emitters
with relatively low surface density, false matches should not
play any role for this object class.

However, when investigating the classes of
optically normal and narrow emission line galaxies,
the problem of false matches has to be taken into account 
due to the high surface densities of these objects. 
In order to check the quality of our optical X-ray counterpart
identification, we applied various tests. 

First, the derived false match rate will depend on the assumed 
position errors which takes into account the statistical 
and the systematic position error (see Sect.~\ref{target}). 
The difference in X-ray and optical position scaled to a 1$\sigma_{\rm{X}}$
position error is shown in Fig.~\ref{diff-x-opt} a. 
When we assume systematic errors of 2 arcsec, the theoretical Gaussian
distribution (Eq.~\ref{distribution_position_error}) 
\begin{eqnarray} 
\label{distribution_position_error} 
f(r_{\rm{n}}) = N_{\rm{tot}} \, r_{\rm{n}} e^{-\frac{1}{2}r_{\rm{n}}^2} \,\,,\,\,r_{\rm{n}}=\frac{r_{\rm{X}}-r_{\rm{O}}}{\sigma_{\rm{X}}} 
\end{eqnarray} 
peaks significantly later than the observed distribution of the X-ray  
counterparts. This is not due to the fact that we systematically accepted
spurious counterparts sources, since the type I AGN distribution (shown with dotted
lines in Fig.~\ref{diff-x-opt}) is consistent with the total X-ray
identification distribution. The fact that we have a large number of type I
AGN allows to estimate the
actual systematic position error for our X-ray observation.    
The observed distribution of the type I AGN position 
errors can be well reproduced with a systematic error
 \begin{figure} 
   \centering 
   \resizebox{\hsize}{!}{ 
   \includegraphics[bbllx=95,bblly=320,bburx=536,bbury=722]{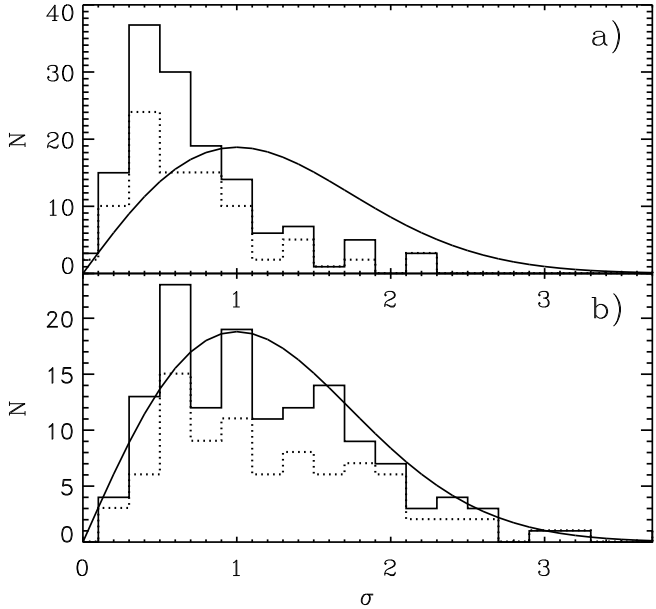}} 
      \caption{\emph{a)} Number of objects identified as X-ray  
               counterparts vs. difference in X-ray and optical position  
               scaled to a 1$\sigma_{\rm{X}}$ position error with a  
               systematic X-ray position error of 2 arcsec. The smooth line  
               represents the theoretical distribution. The distribution  
               of type I AGN, which are a class of well-accepted X-ray
               emitters, is plotted as a dotted line. The solid line  
               characterizes the complete sample including type I AGN, 
               type II AGN, galaxies, and stars; 
               \emph{b)} The  
               1$\sigma_{\rm{X}}$ position error was adjusted to the observed 
               distribution by reducing the systematic X-ray position error  
               from $2\arcsec$ to $0.7\arcsec$. The complete sample agrees
               well with the type I AGN distribution which gives evidences of
               a marginally false match rate.} 
         \label{diff-x-opt} 
\end{figure}  
of $\sigma_{\rm{syst.}} =0.7\arcsec$ (Fig.~\ref{diff-x-opt} b).
Moreover, the distribution of normalized position differences for the total
sample agrees well with the theoretical distribution and with
the type I AGN distribution when we apply $\sigma_{\rm{syst.}} =0.7\arcsec$. 

Considering $\sigma_{\rm{syst.}} =0.7\arcsec$, we calculate the false match rate
by following \cite{sutherland} and \cite{ciliegi}. For every spectroscopically
classified counterpart we determine the likelihood ratio $L$ by
\begin{eqnarray}
L=\frac{Q\,exp(-dist_{\rm{OX}}^2/2)}{2\pi \sigma_{\rm{X}}^2 N(<m_{\rm{R}})}
\end{eqnarray}
where L is the probability of finding the true optical counterpart in this
position with this magnitude, relative to that of finding a similar
chance background object. Hence, the reliability of a counterpart is given by
$P_{\rm true}=L/(1+L)$. 
$Q$ is the probability that the counterpart of the X-ray source is brighter
than the limit of the optical survey and has here been set to $Q=0.5$.
$N(<m_{\rm{R}})$ is the number density of catalogue objects of the relevant class
brighter than the counterpart.
For narrow emission line counterparts we assumed 
that about one third of the field galaxies show line emission detectable
in our spectra (\citealt{kennicutt}). 
We therefore derive the density of background objects
$N(<m_{\rm{R}})$ 
by multiplying the total number of 
objects $<m_{\rm{R}}$ in the optical catalogue by a factor $0.33$.
For normal galaxies a factor of $0.67$ was applied accordingly.

The total number of false matches can then be calculated by adding up all 
probabilities $P_{\rm sp} = 1/(1+L)$
that the counterpart is spurious. Statistically, $\sim$25\% of our normal and 
narrow emission line galaxy are spurious counterparts. 
However, after visually examining objects
with low probabilities $P_{\rm true}$, we find that 
in at least  two cases (422A, 485A) the X-ray position given by the source detection 
software is influenced by blending with another X-ray source and the counterparts
are consistent with the peak of the brighter X-ray source (see Appendix~C, finding chart). 
Taking this into account, we estimate that the false match rate is $\sim$20\%
for the sample of type II AGN. 
For the optically normal galaxies the false match rate calculated with the 
likelihood method is $\sim$2\%.

As a further quality check we plot the position error vs. WFI $R$-band
magnitude. False matches would be recognizable due to different 
distributions of the type I and type II counterparts in this diagram.
However, Fig.~\ref{rmag_dist} shows that all object classes occupy 
the same regions in the diagram.
Hence, the diagram does not indicate any serious contamination by false matches even 
for faint counterparts.
\begin{figure} 
  \centering 
  \resizebox{\hsize}{!}{ 
   \includegraphics[bbllx=95,bblly=372,bburx=535,bbury=695]{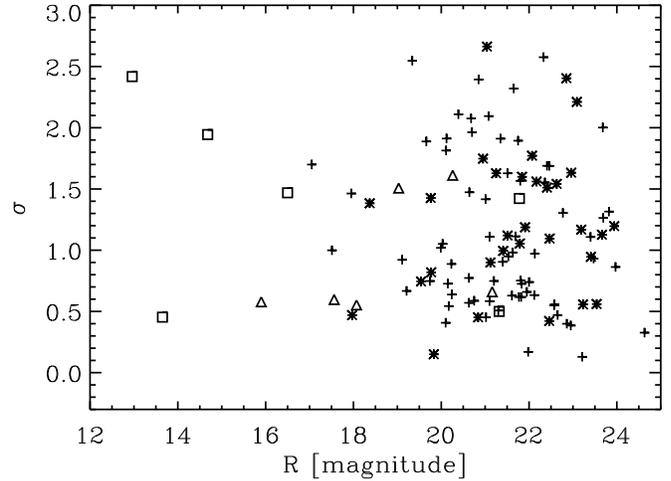}}
      \caption{WFI $R$-band magnitude vs. difference in X-ray and optical
        position   
               scaled to a 1$\sigma_{\rm{X}}$ position error with a  
               systematic X-ray position error of 0.7 arcsec. Crosses mark
               type I AGN, asterisks type II AGN,  
               triangle galaxies, rectangles stars.}
         \label{rmag_dist} 
\end{figure}

\subsection{Classification summary} 
In total, we spectroscopically classified 140 of the optically identified
X-ray sources. Details are shown in 
Table~\ref{tab:fors}.  
\begin{table}
\caption[]{Origin of optical counterparts to XMM sources} 
\label{tab:fors} 
\begin{tabular}{lrr}\hline 
\multicolumn{2}{l}{obtained optical spectra for X-ray sources} & 
207\\\hline\hline 
\multicolumn{2}{l}{positive literature classifications by Marano} & \\ 
\multicolumn{2}{l}{et al. (1988) and \cite{zitelli} (partly reobserved)} & 30\\\hline 
\multicolumn{2}{l}{new optical counterparts of ROSAT sources} &  \\ 
\multicolumn{2}{l}{\cite{zamorani} (partly reobserved)} & 14 \\\hline 
\multicolumn{2}{l}{new positive spectroscopic classifications of XMM sources} & 96\\\hline\hline 
\multicolumn{2}{l}{total positive spectroscopic classifications of XMM sources} & 140\\\hline 
\end{tabular} 
\end{table} 
Like in other deep surveys, e.g. \cite{hasinger1}, the majority of  
X-ray counterparts ($\sim$90\%) is related to the accretion on supermassive  
black holes (type I and type II AGN). 
Furthermore, we classified a few galaxies and stars as optical counterparts 
(see Table~\ref{tab:fors3}).  
\begin{table}
\caption[]{X-ray counterpart distribution} 
\label{tab:fors3} 
\begin{tabular}{lrr}\hline 
 object class & total number &  percentage\\ 
\hline 
 broad emission line objects (B) &89& 63\%\\ 
 narrow emission line objects (N)&36& 26\%\\ 
 galaxies (G)&6 & 4\%\\ 
 stars (S)&9& 7\% \\\hline 
\end{tabular} 
\end{table}

\section{Properties of a core sample of the XMM-Newton Marano survey\label{section4}} 

Our survey suffers somewhat from incomplete optical coverage of the X-ray 
survey area and, hence, a low identification rate over the whole area. 
In order to reach conclusions for the survey of a certain statistical 
significance we constrain our survey area and sample size to the central
0.28 deg$^2$ with a pn-exposure $>$6 ksec and significance of detection 
of individual sources with $ML >10$. We refer to this as the 'core sample'.
See Fig. \ref{exposure_map2} for the definition of the core region. Note that 
the maximum exposure is 35.5 ksec for the PN camera and 78 ksec for each of
the MOS cameras.

The core sample contains 170 X-ray sources ($ML>10$).
No optical data are available for six out of the 170 X-ray
sources which fall outside the WFI $R$-band image and are also not covered
by VLT pre-images. Further six X-ray sources from the core sample have no 
optical detection in the WFI $R$-band image in a 3$\sigma_{\rm{X}}$ position
error circle ($\sigma_{\rm{syst.}} =0.7\arcsec$). 
122 X-ray sources are new detections, while 35 are spectroscopically
classified ROSAT X-ray 
sources, and 13 are optically unidentified ROSAT sources. Out of the 140 
spectroscopically classified objects in the Marano Field, 110 are
associated to X-ray sources of the core sample. A summary of the
properties of the core sample is given in
Table~\ref{tab:core}. 

Fig.~\ref{mag_r_core} shows $R$-band magnitude
histograms of the core sample X-ray sources that were spectroscopically
classified and only optically identified respectively. 
The ($f_{\rm{X}}$,$ML$)-distribution of X-ray sources in the core region 
is given in Fig.~\ref{defining_sample}.

The identification ratio
of the core sample is 65\%.
In the next subsections we always refer to this sample. 

\begin{table}
\caption[]{Properties of the core sample} 
\label{tab:core} 
\begin{tabular}{lrr}\hline 
 X-ray sources ($t_{\rm{PN}}>6$\,ksec, $ML >10)$ & & 170\\ 
\hline 
 optically identified X-ray counterparts & & 158\\ 
 spectroscopically classified optical counterparts&  & 110\\ 
 \hline 
\end{tabular} 
\end{table} 

\begin{figure}
   \centering 
   \resizebox{\hsize}{!}{ 
   \includegraphics[bbllx=95,bblly=372,bburx=545,bbury=695]{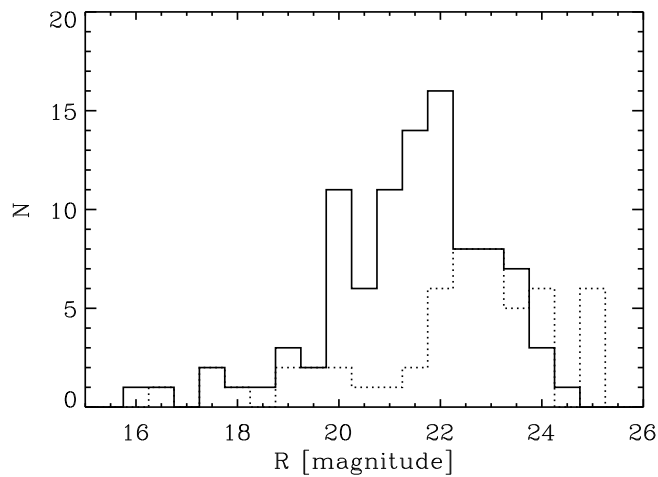}} 
      \caption{WFI $R$-band magnitude histogram for the core sample in the
   XMM-Newton Marano Field survey. The solid line represents the $R$-magnitude
   distribution of the spectroscopically classified X-ray counterparts. The 
   dashed line shows the distribution of the optically identified but not
   spectroscopically classified X-ray counterparts. 
   Six X-ray sources, which have no counterpart within a 3$\sigma_{\rm{X}}$ 
   error circle, are assigned to the bin at  $R$=25.0.}
     \label{mag_r_core} 
\end{figure}

\begin{figure}
   \centering 
   \resizebox{\hsize}{!}{ 
   \includegraphics[bbllx=60,bblly=374,bburx=555,bbury=701]{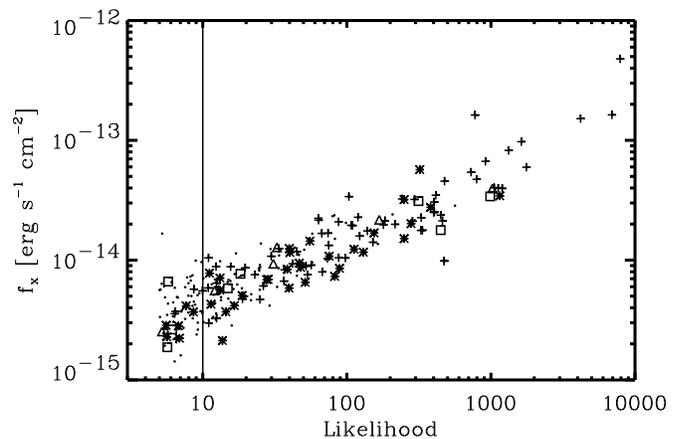}} 
      \caption{Likelihood vs. 0.2-10 keV X-ray-flux of the 252 central X-ray 
       sources (pn-detector exposure time $\ge$6 ksec).  
            Crosses mark type I AGN, asterisks type II AGN, 
               triangle galaxies, rectangles stars. Dots indicate 
               X-ray sources with no spectroscopic classification.  
               The vertical line indicates $ML = 10$.} 
         \label{defining_sample} 
\end{figure} 
 
\subsection{X-ray properties \label{argument}} 
Characterising X-ray sources only by the X-ray properties  
can be used to reveal and study the existence of  
different X-ray populations and their features.   
Hardness ratios (see Sect.~\ref{xray}) are the simplest tool to determine the  
spectral energy distribution in the X-ray regime. 
In Fig.~\ref{fig:hr3hr2} ($HR2$ vs. $HR3$) we only plot those 
X-ray sources in an X-ray  
color-color diagram which have $HR2$ error $\sigma_{\rm{HR2}} \le 0.3$.
The different spectroscopically classified classes occupy different regions 
in this X-ray color-color diagram. 
A noticable separation in type I and type II AGN can be recognized. 
This result is in agreement with \cite{mainieri} and \cite{caccianiga}. 
Type II AGN spread over a much broader $HR2$ ($0.9>HR2>-0.6$).  
The lowest $HR2$ values for type II AGN overlap with the highest $HR2$ values 
for type I AGN. However, we do not see a large fraction of type II AGN 
occupying the $HR2$ range typical for type I AGN, as was reported by 
\cite{dellaceca} for the XMM-Newton bright serendipitous survey.
Fig.~\ref{fig:hr3hr2} shows that $HR2$ is a good indicator for intrinsic
absorption. 
The X-ray sources with $HR3 = -1.0$ correspond to
non-detections in the 4.5-7.5 keV band.  
The objects with $HR3 > 0.4$ and soft $HR2$ values 
have $HR3$ errors $ \geq 0.3$ (small plot symbols).
Therefore, it is likely that their deviation from the typical location
of sources in the $HR2$ vs. $HR3$ diagram is caused by statistical fluctuations. 
 \begin{figure}
   \centering 
   \resizebox{\hsize}{!}{ 
   \includegraphics[bbllx=78,bblly=374,bburx=533,bbury=695]{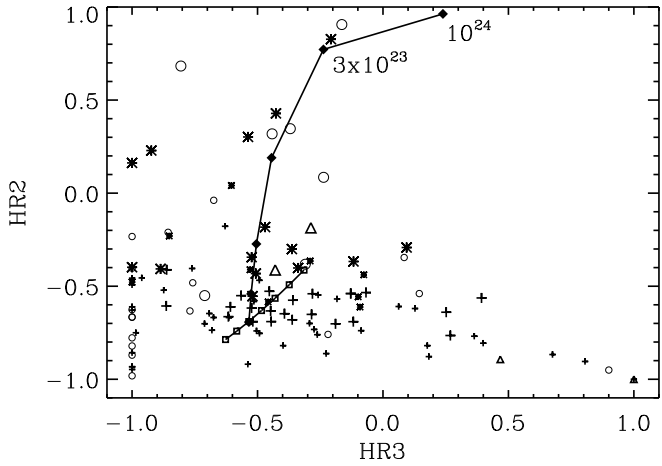}} 
      \caption{X-ray diagnostic diagram based on hardness ratios. $HR2$ is 
   calculated by  
   using the 0.5-2.0 keV \& 2.0-4.5 keV band, $HR3$ 2.0-4.5 keV \& 4.5-7.5 keV.  
   Plotted 
   are only X-ray sources that have $\sigma_{\rm{HR2}} < 0.3$. Labels: 
   crosses - 63 type I AGN, asterisks - 24 type II AGN, triangles - 4 galaxies, 
   open circles - 24 unidentified X-ray sources with $ML > 10$. 
   Large symbols represent objects with $\sigma_{\rm{HR3}} < 0.3$, 
   small symbols $\sigma_{\rm{HR3}} \ge 0.3$.
   The solid line with box symbols (at $HR2\sim HR3\sim-0.5$) represents an unabsorbed powerlaw X-ray spectrum 
   (corrected for galactic absorption) with different photon index 
    $\Gamma=2.4-1.2$  in steps of 0.2 (squares).  Different hydrogen
   column densities ($N_{\rm{H}}$/cm$^{-2}= 10^{20},10^{22},3\times
   10^{22},10^{23},3\times 10^{23},10^{24}$; see Sect.~\ref{nh}) are plotted 
   as a solid line with diamonds for $\Gamma=2$ and $z=1$. } 
         \label{fig:hr3hr2} 
\end{figure} 

The small number of  optically normal galaxies span a large range in $HR2$.
Two have $HR2$ values that belong to the softest in the whole core sample.
The other two have X-ray spectra similar to those of  
soft type II AGN or very hard type I AGN.

In addition to the spectroscopically classified objects we plot also unidentified objects 
from the core region. Out of the total 60 unidentified X-ray 
sources, only 24 meet the selection criterion of $\sigma_{\rm{HR2}} \le 0.3$.
Based on the rather clear separation between type I and II AGN one may assign
a likely classification to the yet unidentified sources. Among the 24 
unidentified sources with reliable X-ray colors the numbers of type I and
type II AGN candidates appear to be similar.
 
\subsection{Redshift distribution \label{AGN_discussion}} 

\begin{figure}
   \centering 
   \resizebox{\hsize}{!}{ 
   \includegraphics[bbllx=88,bblly=320,bburx=536,bbury=871]{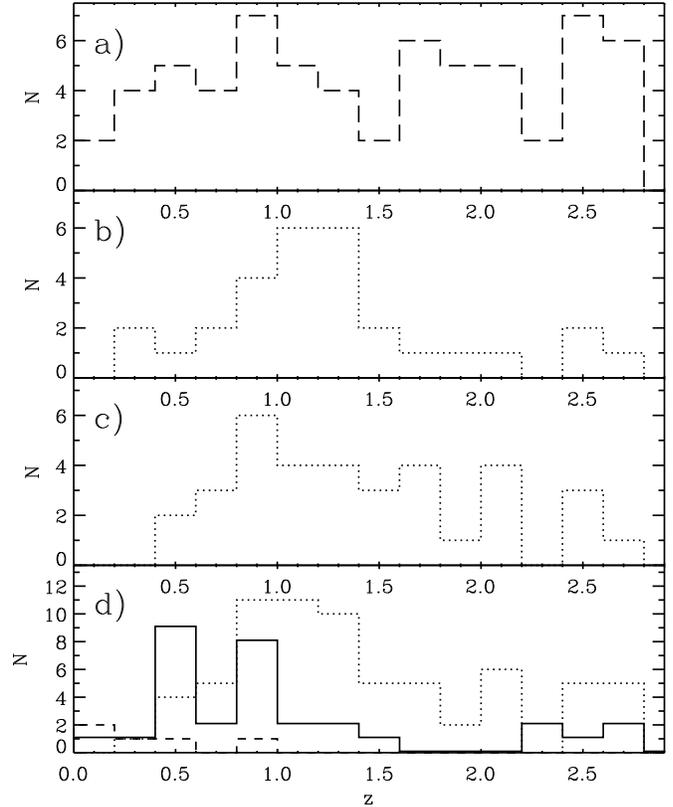}} 
      \caption{Redshift distribution of AGN in the Marano Field. {\emph a)} 
               complete sample of 64 optically selected quasars 
               ($B_{\mathrm{J}} \leq 22.0$) by \cite{zitelli}; {\emph b)}  
               ROSAT X-ray sources (29 type I AGN by 
               \cite{zamorani}); {\emph c)} 35 type I AGN newly detected by 
               XMM-Newton (core region); {\emph d)} 
               Redshift distribution of the core sample.   
               Dotted line: all 70 XMM-Newton detected type I AGN,  
               solid line: 31 XMM-Newton detected type II AGN,  
               dashed line: 5 XMM-Newton detected optically normal galaxies.} 
         \label{N_over_z_4panels} 
\end{figure} 

Previous extensive studies of AGN in the Marano Field  
enable us to compare these samples with our XMM-Newton 
detections. 
The optical survey by \cite{zitelli} covers $\sim$0.7$~{\rm deg}^2$.  
The selected quasars, which are all of type I, show an almost flat 
distribution in redshift (Fig.~\ref{N_over_z_4panels} a) up to $z=2.8$.
The ROSAT $0.2~{\rm deg}^2$ survey in the field (\citealt{zamorani}) 
recovered most of the optically selected quasars at redshifts 
up to $z\sim 1.4$ (Fig. \ref{N_over_z_4panels} b).
The newly detected ROSAT AGN, with few exceptions, are type I AGN, 
which is expected due to the limited capability of ROSAT to detect absorbed 
sources.
 
In the $0.28~{\rm deg}^2$ core region of the XMM-survey we have detected 
23 of the 29 broad emission line quasars of the optically selected sample 
of \cite{zitelli}. The detection rate of optically selected quasars 
remains constantly high over all redshifts.

However, looking at the type I AGN newly discovered with XMM-Newton,
(Fig. \ref{N_over_z_4panels} c) it is apparent that the X-ray selection 
tends to detect quasars at lower redshifts than the optical surveys. This is 
particularly obvious from the redshift distribution of the ROSAT detected 
quasars, representing the brightest X-ray sources in the field.
But also the mean redshift $z \sim 1.3$ of the XMM-Newton detected type I AGN is
lower than that of the optically selected sample $z \sim 1.5$, despite the fact
that the XMM-Newton-observations are deeper in terms of the surface density of quasars
than the optical survey.
Possible reasons for these differences are discussed in Sect.~\ref{section5}.
 
The redshift distribution according to object 
class of the XMM-Newton detected core sample
is given in Fig.~\ref{N_over_z_4panels} d.
We find that almost half of the new XMM-Newton
sources are classified as type II AGN with redshifts mostly below 1.0.  
Type I AGN extend over a wide range of redshifts  
with a maximum at $z\sim0.8-1.4$. Type II AGN are comparable to type I AGN 
in number density at low redshifts, but are mostly found below $z=1$. 
Five type II quasars at $z > 2.2$ have been identified. 
Optically normal galaxies without emission lines are found at $z < 0.9$.
 
\subsection{Observed X-ray luminosity} 

For the X-ray sources with measured redshifts X-ray luminosities can be 
computed. 
The coverage of the survey in \mbox{redshift -- X-ray} luminosity space 
is plotted in Fig.~\ref{lx_z}. 
\cite{szokoly} showed that the different object classes
identified in X-ray surveys occupy different regions in 
a diagram of hardness ratios versus observed X-ray luminosities. 
In Fig.~\ref{fig:lxhr2} 
we follow a scheme similar to that adopted by \cite{szokoly} 
and consider objects 
with an X-ray luminosity log$(L_{\rm{X}}) \ge 44.0$ as quasars (QSOs). 
The majority of type I AGN ($\sim$70~\%) are actually type I QSOs. Most of 
type II AGN ($\sim$71~\%) are low X-ray luminosity objects. 
Only type II objects with redshifts $z \ge 2.0$ (marked in grey in 
Fig.~\ref{fig:lxhr2}) have X-ray luminosities of type II QSOs.    
  
\begin{figure}
   \centering 
   \resizebox{\hsize}{!}{ 
   \includegraphics[bbllx=94,bblly=374,bburx=543,bbury=695]{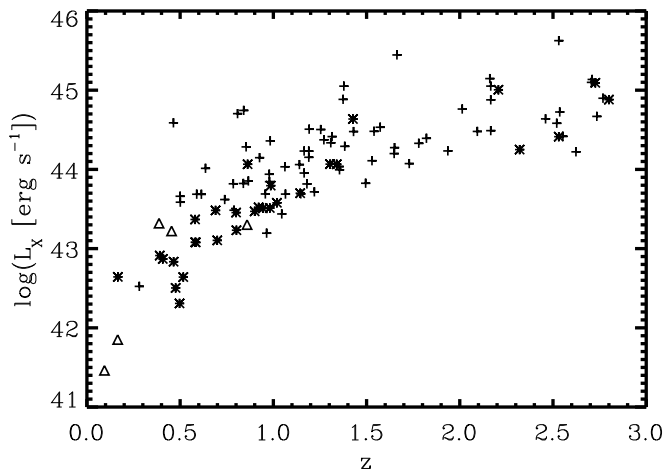}} 
      \caption{Observed X-ray luminosity (0.2-10 keV) vs. redshift.  
      Labels: crosses - 70 type I AGN, asterisks - 30 type II AGN,  
        triangles - 5 galaxies.} 
         \label{lx_z} 
\end{figure}

\begin{figure}
   \centering 
   \resizebox{\hsize}{!}{ 
   \includegraphics[bbllx=78,bblly=368,bburx=539,bbury=695]{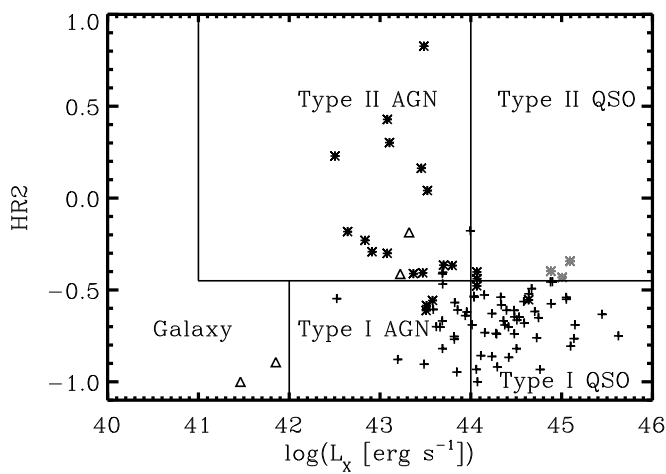}} 
      \caption{Observed X-ray luminosity (0.2-10 keV) vs. hardness ratio. $HR2$ 
   is calculated by using 0.5-2.0 keV \& 2.0-4.5 keV bands. Plotted 
   are only X-ray sources that have $\sigma_{\rm{HR2}} < 0.3$. Labels: 
   crosses - 63 type I AGN, asterisks - 24 type II AGN (grey marked have $z \ge 
   2.0$), triangles - 4 galaxies. The diagram separation into different regions
   is based on \cite{szokoly} with an adjusted threshold in the $HR2$-value.} 
         \label{fig:lxhr2} 
\end{figure} 

High-redshift type II QSOs with intrinsic absorption are found to be 
indistinguishable from non-absorbed type I AGN on the basis of their X-ray
spectral hardness ratios, since the absorbed part of the spectrum is shifted
out of the observable spectral window towards lower energies. This explains
the emptiness of the upper right corner of Fig.~\ref{fig:lxhr2} labeled 
'Type II QSO', 
a classification which applies to low redshift objects only.

Optically normal galaxies vary clearly in X-ray luminosity. The $HR2$-soft objects have
very low X-ray luminosities.
The two $HR2$-hard normal galaxies  
are found in the same region as the softest type II AGN, but are harder than  type I AGN.  

\cite{szokoly} use $HR2=-0.2$ as a threshold for the separation of 
type I and type II objects. Assuming a $\Gamma=2$, this value corresponds
  to a hydrogen column density \mbox{$N_{\rm{H}}$/cm$^{-2}=10^{22}, 10^{23}$} for 
$z= 0.25, 2.1$, respectively. For their CHANDRA observation they computed 
$L_{\rm{X}}$ in the \mbox{0.5-10 keV} band and their hardness ratio was based on the
\mbox{0.5-2 keV \& 2-10 keV} bands. Because their definition of the X-ray bands differs from 
our study, the
majority of their objects have higher hardness ratios (compared to our $HR2$). 
Furthermore, the XMM-Newton 
pn-detector, that was used for calculating the hardness ratios, has a higher  
efficiency in the 0.5-2 keV band compared to the CHANDRA detector. 
Therefore, we lowered the $HR2$ threshold from $HR2 = -0.2$ in  
\cite{szokoly} to $HR2 = -0.45$. 
Our threshold corresponds to
$N_{\rm{H}}$/cm$^{-2}=0.54\times 10^{22}, 1.7\times 10^{22}, 5.4\times
10^{22}$ for $\Gamma=2$ and $z= 0.25, 1, 2.1$. This is about two times 
lower than the $N_{\rm H}$ cutoff
\cite{szokoly} are using.

\subsection{N$_{\rm{H}}$ column densities and corrected X-ray luminosities \label{nh}} 
 
The  hardness ratio diagram (Fig.~\ref{fig:hr3hr2}) supports 
the view that the majority of type II AGN and a small fraction of 
type I AGN are obscured sources. Hence, the observed X-ray luminosity 
does not represent the intrinsic object X-ray luminosity.    

The significant deficit of soft photons as compared to a power law spectrum 
reflects the existence of an absorbing component that is expressed by the 
hydrogen column density $N_{\rm{H}}$. 
For most of the sources the number of detected counts is not sufficient 
to extract a spectrum and fit a power law model  with $N_{\rm{H}}$
and the photon index $\Gamma$ as free parameters.
We, therefore, applied a technique which uses the measured hardness ratios to calculate the $N_{\rm{H}}$
value for each source with a set of fixed power law indices.
 \cite{mainieri} 
found a mean value of $<\Gamma>\simeq 2$ for 61 
type I and type II AGN in the Lockman Hole. The majority of type I and II AGN
are found in the range of $\Gamma\simeq 1.7-2.3$. The finding is confirmed 
by \cite{mateos}. They find $<\Gamma>= 1.92$ with a $\sigma =0.28$.
Therefore, we use the observed pn-, mos1-, and mos2-hardness-ratios 
(0.2-0.5 keV \& 0.5-2.0 keV \& 2.0-4.5 keV \& 4.5-7.5 keV) and performed three
runs to determine $N_{\rm{H}}$ with the values $\Gamma= 1.7, 2.0, 2.3$ 
for all objects.   

First of all, we computed a grid of model hardness ratios for  
all EPIC instruments with {\tt Xspec}  
(using the models {\tt wabs}, {\tt zwabs}, and {\tt powerlaw}).  
As input the galactic absorption in the line of sight in the field
with $N_{\rm{H}} = 2.7 \times 10^{20}$ cm$^{-2}$, the redshift 
of the object and a grid of hydrogen column densities 
($N_{\rm{H}}=0,\, N_{\rm{H}} = 10^{20+0.04\,a}\ \mathrm{cm^{-2}},\ \ a \in 
N_{\rm{0}},\ \ a=0..100$) is used. 
We then computed the $\chi^2$ values for the deviations of the
measured  hardness ratios and their model values, summed over 
the  three instruments and the three hardness ratios $HR1$, $HR2$, $HR3$.
The procedure was applied only to 
those sources which had at least five out of nine hardness ratios 
with $\sigma HR_{\rm{obs,i}} < 0.7$.\\

The $N_{\rm{H}}$ 
of a source is determined by finding the minimum of
\[
\chi^2(N_{\rm{H}})=\sum_i\frac{(HR_{\rm{obs,i}}-HR_{\rm{model,i}}(N_{\rm{H}}))^2}{\sigma
  HR_{\rm{obs,i}}^2}\ \ .\]
The models  are based on the photon index  $\Gamma= 2$.

The statistical 1$\sigma$ errors were derived from the range in $N_{\rm{H}}$,
where $\chi^2 < \chi^2_{\rm min}+1$ (\citealt{lampton}). 
For the error calculation we also took into account an intrinsic scatter
in the photon indices. Its contribution to the  $N_{\rm{H}}$ error  
was measured by finding the
minimum $\chi^2$ for each source in the grids calculated using
 $\Gamma= 1.7$ and  $\Gamma= 2.3$.
The resulting systematic errors were quadratically added to the statistical errors.
The  $N_{\rm{H}}$ values and the total errors are given in Table~\ref{opt_tab}.
Due to the uncertainties in the  $N_{\rm{H}}$ determination we regard all values
with $N_{\rm{H}} \leq 10^{21}~\mathrm{cm^{-2}}$ as consistent with unabsorbed 
spectra.   

We tested our procedure by performing an individual {\tt Xspec} $N_{\rm{H}}$-fit 
for the brightest type II AGN, which has sufficient X-ray data quality.
The hardness ratios of 32A (z=2.727)
indicate the highest absorption among the brightest type II AGN. 
A spectral fit with  
{\tt wabs}, {\tt zwabs}, {\tt powerlaw}, and $\Gamma$ as a free parameter determined  
\mbox{$N_{\rm{H_{32A}}}=(10.0 \pm 3.5) \times 10^{22}~\mathrm{cm^{-2}}$} and
\mbox{$\Gamma_{\rm{32A}}=(1.7 \pm 0.3$).} 
The best fit $N_{\rm{H}}$, for a fixed value of $\Gamma= 1.7$, with the  
hardness-ratio $\chi^2$-minimum fit is
\mbox{$N_{\rm{H_{32A}}}=(7.5~_{-2.8}^{+3.4}) \times 10^{22}~\mathrm{cm^{-2}}$.}

A spectral fit with a fixed $\Gamma_{\rm{32A}}=2.0$ results in 
\mbox{$N_{\rm{H_{32A}}}=(13.4 \pm 2.5) \times 10^{22}~\mathrm{cm^{-2}}$.} 
The given $N_{\rm{H}}$ in Table~\ref{opt_tab} was performed with the hardness-ratio  
$\chi^2$-minimum fit and finds 
$N_{\rm{H_{32A}}}=(14.5~_{-8.2}^{+7.9}) \times 10^{22}~\mathrm{cm^{-2}}$. Hence, both
fit methods give comparable results, at least for high SNR sources.

Fig.~\ref{nh_hist} shows the computed hydrogen column densities  
for type II AGN and optically normal galaxies. 
\begin{figure}
   \centering 
   \resizebox{\hsize}{!}{ 
   \includegraphics[bbllx=94,bblly=370,bburx=550,bbury=695]{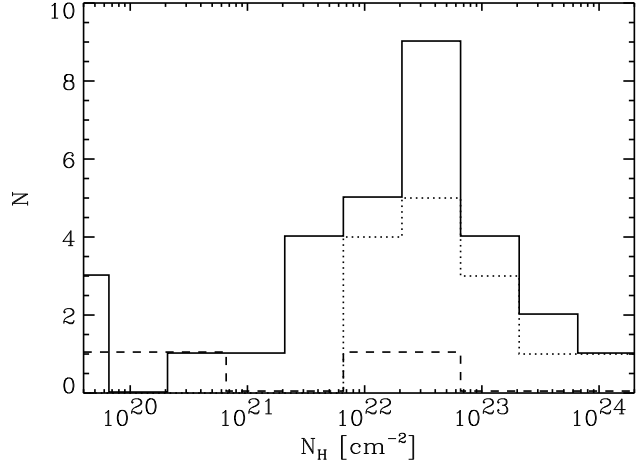}} 
      \caption{Calculated intrinsic hydrogen column density histogram for the  
        central Marano Field. The solid line shows the distribution of 
        type II AGN that have reasonable fits in hydrogen column density.
        Objects with
        $N_{\rm{H}}=0$ are set to $N_{\rm{H}}=5 \times 10^{19}~\mathrm{cm^{-2}}$. 
        The dotted line represent only type II AGN that show intrinsic absorption
        with a significance higher than $2 \sigma$  in hydrogen column density. 
        Optically normal galaxies are plotted as a dashed line. 
        Dotted line - 14 type 
        I AGN; solid line - 30 type II AGN; dashed line - 5 galaxies.} 
   \label{nh_hist} 
\end{figure} 
As expected, the majority of type II AGN shows absorption. 
However, 13\% of the type II objects (51A, 132A, 133A, 607A) have absorbing
column densities $N_{\rm{H}} < 10^{21}~\mathrm{cm}^{-2}$. 
All of the type II AGN discovered by \cite{mainieri} 
in the Lockman Hole have absorbed X-ray spectra. 
The values for optically normal galaxies range from unabsorbed to moderately
absorbed. Two of the unabsorbed galaxies have low X-ray luminosities 
($L_{\rm X} < 10^{42}\,{\rm erg\,s^{-1}}$). 

The computed hydrogen column density of type II AGN object 145A is the highest
in our sample ($N_{\rm{H}} = 10^{24}~\mathrm{cm}^{-2}$). Since this 
corresponds to the highest value in the model grid, no reliable error estimate
can be given for this object. However, the extremely high column density for
this object is confirmed by the fact that it is one of the few sources
detected in the \mbox{$4.5-7.5\,{\rm keV}$} EPIC images, but it is not visible in the
softer bands. 
 
Based on the  hydrogen column densities we calculated the unabsorbed  
(intrinsic) X-ray luminosities by computing a correction factor for the
observed X-ray flux.
Fig.~\ref{3Lx_plot} shows the comparison of the 
absorbed (a) and unabsorbed (b) X-ray luminosity distributions.
Even after correcting the X-ray luminosity, type II AGN  
have lower median X-ray luminosities than type I AGN, although
type I and II AGN cover the same X-ray luminosity range.
In contrary to the almost flat distribution of type II AGN with a 
median of $L_{\rm{X}}\sim 10^{44}\,\mathrm{erg~s^{-1}}$, type I AGN show 
a significant peak at $L_{\rm{X}}\sim 10^{44.4}\,\mathrm{erg~s^{-1}}$
in observed and intrinsic X-ray luminosity.
Type I AGN 5A exhibits the highest X-ray luminosity in our sample. 

The hydrogen column density as a function of the intrinsic X-ray luminosity
is shown in Fig.~\ref{3Lx_plot} c). \cite{mainieri} suggest to label the
region defined by $N_{\rm{H}} > 10^{22}~\mathrm{cm}^{-2}$ and $L_{\rm{X}} > 10^{44}\,\mathrm{erg~s^{-1}}$
as ``type II QSO region''. They proposed this classification 
based on 61 AGN identified from XMM-Newton sources in the Lockman Hole.
Following this classification our sample includes 10 type II QSOs.

In our case this region of the plot is also populated by nine type I QSOs,
which formally have intrinsic hydrogen column densities 
$N_{\rm{H}} > 10^{22}~\mathrm{cm}^{-2}$. However, most of these 
detections of intrinsic $N_{\rm{H}}$ have low significance, only
four of all type I QSOs show absorption at the $2 \sigma$ level
(see Table~\ref{absorbed}).  

For about half of the type II AGN significant intrinsic absorption
 ($N_{\rm{H}} > 2\times \Delta N_{\rm{H}} $)  was measured, no
dependence of the absorbed fraction on luminosity is evident
(Table~\ref{absorbed}). 

\begin{figure}[t] 
   \centering 
   \resizebox{\hsize}{!}{ 
   \includegraphics[bbllx=70,bblly=340,bburx=545,bbury=871]{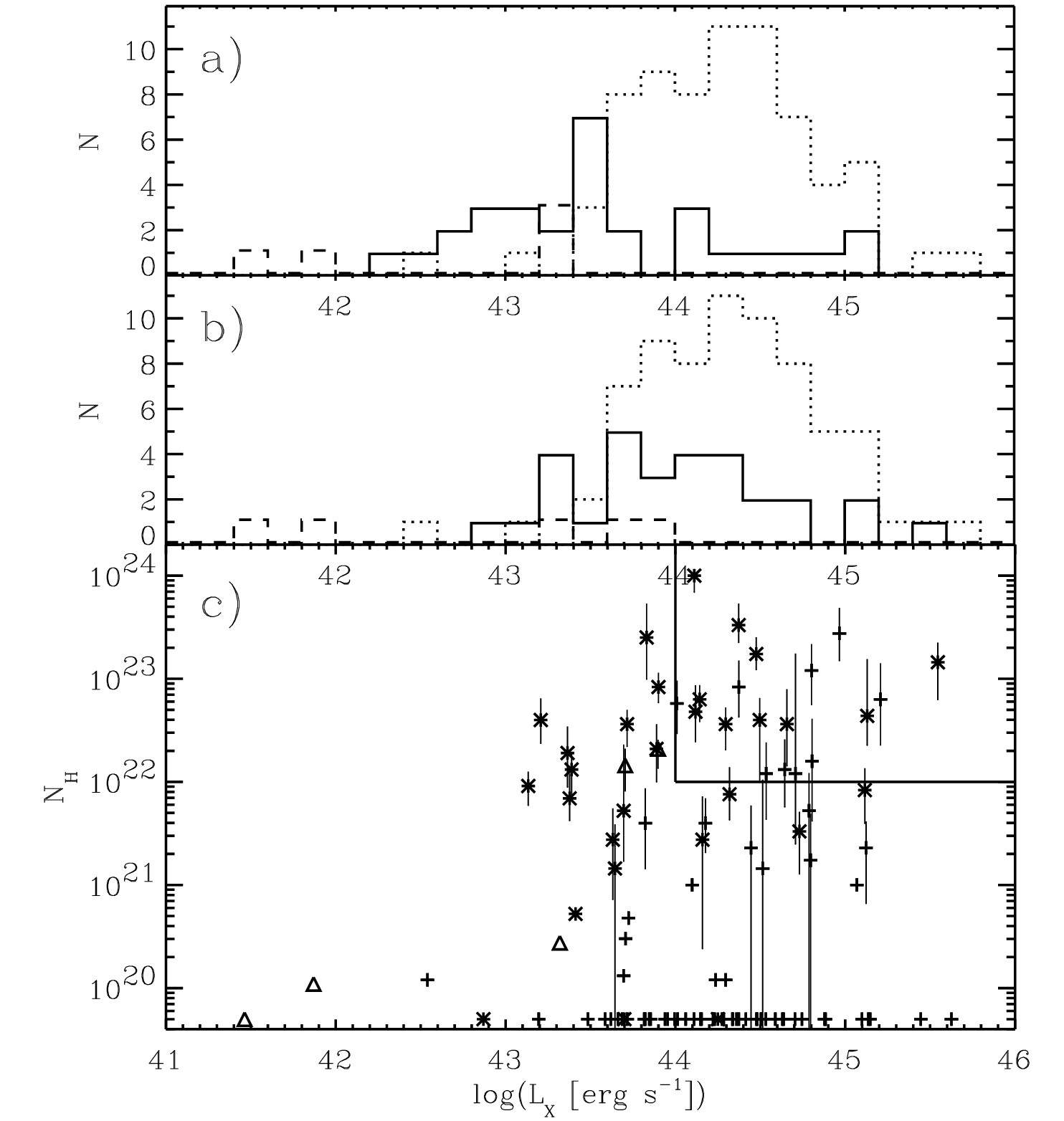}\vspace*{0.5cm}} 
      \caption{\emph{a)} Histogram of the absorbed (observed)
                X-ray luminosity (0.2-10 keV) for the core sample. 
               Dotted line - 70 type 
        I AGN, solid line - 30 type II AGN, dashed line - 5 galaxies;         
        \emph{b)} Histogram of the X-ray luminosity (0.2-10 keV) corrected
        for intrinsic absorption (core sample). Same notation as in a);
        \emph{c)} Corrected X-ray luminosity \mbox{(0.2-10 keV)} vs. hydrogen 
        column density (core sample). Crosses - type I AGN, 
        asterisks - type II AGN, triangles - galaxies. 
        The upper right corner is suggested to be defined as ``Type II QSO region'' by
        \cite{mainieri}. Objects with
        $N_{\rm{H}}=0$ are set to $N_{\rm{H}}=5 \times 10^{19}~\mathrm{cm^{-2}}$.} 
         \label{3Lx_plot} 
\end{figure} 

\begin{table}[t] 
\begin{center} 
\caption{Fractions of absorbed  sources  
         in different X-ray luminosity bins based on Fig.~\ref{3Lx_plot} c. Only  objects with $N_{\rm{H}}> 2\times \Delta N_{\rm{H}}$ are selected.} 
 \begin{tabular}{cccc}\hline \label{absorbed}  
 & $ 43 \le \mathrm{log}(L_{\rm{X}}) < 44$ & $ 44 \le \mathrm{log}(L_{\rm{X}}) < 45$ 
 & $\mathrm{log}(L_{\rm{X}}) \ge 45$ \\ \hline 
type I & 0\% (0/19) &10\% (4/42) &0\% (0/8) \\ 
type II & 43\% (6/14) &58\% (7/12) &33\% (1/3) \\\hline 
 \end{tabular} 
\end{center}  
\end{table}

\subsection{X-ray to optical flux ratios} 
The ratio between X-ray flux and optical flux ($f_{\rm{X}}/f_{\rm{opt}}$)
is used in former deep X-ray surveys to characterize the  
different X-ray emitting classes (\citealt{szokoly, mainieri}). 

In order to calculate   $f_{\rm{X}}/f_{\rm{opt}}$ values we derived
optical fluxes in a band centered at $7000\,\AA$
and width $1000\,\AA$ using the equation $f_{\rm{opt}}=10^{-0.4 R - 5.759}$
(\citealt{zombeck}).
As X-ray fluxes we used the 0.2-10$\,{\rm keV}$ values (see section \ref{xray}).
  
The distribution of our core sample in the $(R-f_{\rm{X}}/f_{\rm{opt}})$-plane 
is illustrated in Fig.~\ref{fx_R}. In general, type I AGN show higher X-ray 
fluxes and are brighter in the $R$-band. Type II AGN are found at lower X-ray 
fluxes and have fainter $R$-band counterparts. 
Two of the optically normal galaxies have X-ray to optical flux ratios similar to type I and type II AGN.
Another two are among the objects with lowest  X-ray to optical flux ratios in the sample.

\begin{figure}
   \centering 
   \resizebox{\hsize}{!}{ 
   \includegraphics[bbllx=60,bblly=370,bburx=533,bbury=702]{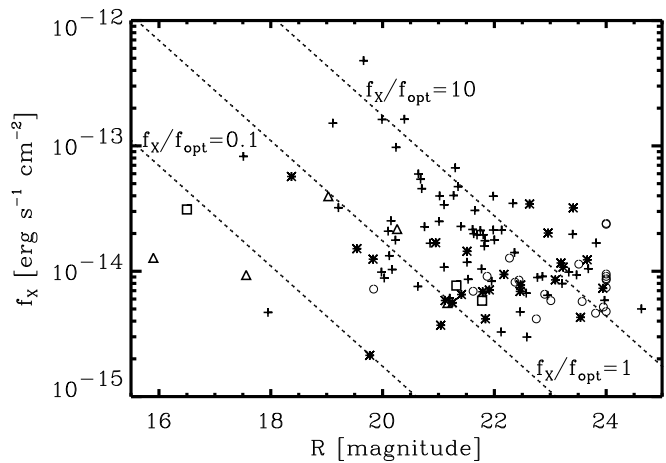}} 
      \caption{Observed X-ray flux (0.2-10 keV) vs. $R$-band magnitude. 
       Dashed  lines indicate  
       the $f_{\rm{X}}/f_{\rm{opt}}$ values 0.1, 1.0, and 10.  
       Labels: crosses - 63 type I AGN, asterisks  
       - 25 type II AGN, triangles - 5 galaxies, rectangles - 3 stars, 
        circles - 21 spectroscopically unidentified 
       sources ($R$-band lower limits are plotted as circles at $R$=24.0).
       Note that for a few objects of the core sample no WFI $R$-magnitudes
       were available and therefore these are not plotted here.}  
         \label{fx_R} 
\end{figure} 

Our type I and II AGN show a large variety in X-ray flux and $R$-band magnitude. 
They are detected at X-ray fluxes from 
$f_{\rm{X}}\sim 2 \times 10^{-15}$ to $5 \times 10^{-13}$\,erg\ cm$^{-2}$\,s$^{-1}$ and in $R$-band
magnitudes from $R \sim 18$ down to the detection limit of 
$R \sim 24$.  
There is one type I AGN (191A) at an unusually low value $f_{\rm{X}}/f_{\rm{opt}}=0.04$, 
a factor 100 below the mean value of type I AGN.
The highest ratio in X-ray to optical 
flux is found for the type II AGN object 39A with $f_{\rm{X}}/f_{\rm{opt}}=42.5$.

Most of the stars are found at star-typical 
$f_{\rm{X}}/f_{\rm{opt}} < 0.05$ with $R < 17$. 
Nevertheless, we also detected two M-stars with 
$f_{\rm{X}}/f_{\rm{opt}} \sim 1$ and $R \sim 21.5$.  

Unidentified sources with WFI $R$-band data are plotted as circles 
in Fig.~\ref{r-k_r}. An analysis of the WFI $R$-band catalogue  
showed that we are significantly losing completeness at 
\mbox{$R = 24$.}  
Hence, unidentified objects that have no detection in the $R$-band catalogue  
are plotted at the mentioned threshold and represent upper limits.

\subsection{Optical-to-near-IR colors} 

 Fig.~\ref{r-k_r} shows that type II AGN tend to be fainter
and redder in the optical window than type I AGN.
This figure shows a general trend for both type I and type II AGN 
to become redder for fainter R-magnitudes. The 
lack of faint blue objects can be explained by the $K$-band detection limit
of $K$=20.0, where we lose completeness (see Fig.~\ref{fig:magK_histogram}).
However, the lack of bright red objects cannot be caused by 
any detection bias.
The type II AGN have redder $R-K$ colors, although with some overlap with the 
reddest type I AGN in the sample. 
These trends can be explained by an increasing contribution
of the host galaxies for fainter type I AGN and type II AGN.
The faint $R$-magnitudes and high $R-K$ values indicate 
higher optical obscurations in type II AGN. This is in agreement with previous 
X-ray studies of these objects that revealed a significantly higher ratio of 
absorbed to unabsorbed objects compared to type I AGN.

In our survey type I and type II  objects do not separate as clearly  as seen  
in a similar plot in \cite{mainieri}. Therefore, we checked whether the reddest
and faintest type I objects are reliably classified. The most extreme type I objects 
are 69A, 84A, and 585A. Their optical spectra clearly show broad emission lines, but their
continua are redder than typical type I spectra. In addition, object 585A shows only weak emission lines.    

\begin{figure}
   \centering 
   \resizebox{\hsize}{!}{ 
   \includegraphics[bbllx=103,bblly=371,bburx=533,bbury=695]{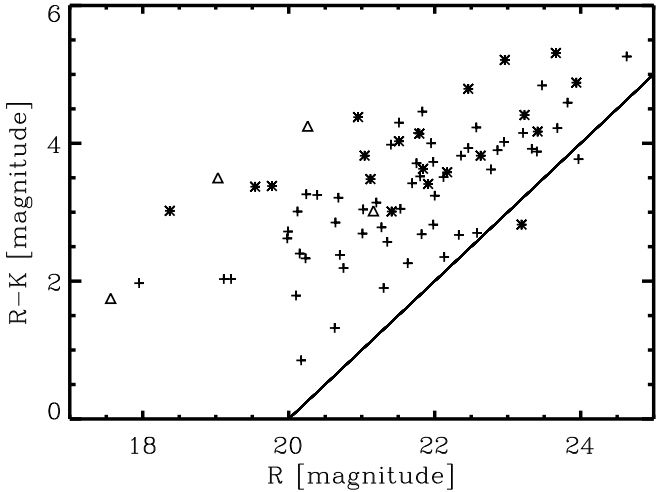}} 
      \caption{Color magnitude diagram of spectroscopically classified X-ray sources.  
        Labels: crosses - 56 type I AGN, asterisks  
       - 20 type II AGN, triangles - 4 galaxies. Only objects with detections in
         both $R$ and $K$ are included. The absence of blue, faint
       objects (below the solid line) is due only to the $K$-magnitude limit.} 
         \label{r-k_r} 
\end{figure} 
 
The optically normal galaxies are brighter in $R$-magnitudes than the typical type II AGN, 
but show similar  
$R-K$. All classified galaxies have $R<21.5$. 
The majority of type II AGN and galaxies have $R-K > 3\,{\mathrm{mag}}$,
consistent with the spectrum of the host galaxy being the dominating component.
Also a considerable fraction of type I AGN have red $R-K$-colors.
These are mostly low luminosity objects, where the host galaxy may also
dominate the optical continuum. 

\section{Additional objects\label{section6}} 
In Table~D.1 (Online Material, 
Appendix~D) we list objects in the Marano Field  
that are not related to X-ray sources. These objects were  
spectroscopically investigated since slit positions on the  
multi-object spectroscopy masks were still available.
The random selection of these additional objects, which follow a similar
$R$-magnitude distribution as the sample of X-ray selected type II objects, 
is useful to investigate the
redshift distribution of the field galaxies in the Marano Field.

\begin{figure}
   \centering 
   \resizebox{\hsize}{!}{ 
   \includegraphics[bbllx=94,bblly=374,bburx=533,bbury=695]{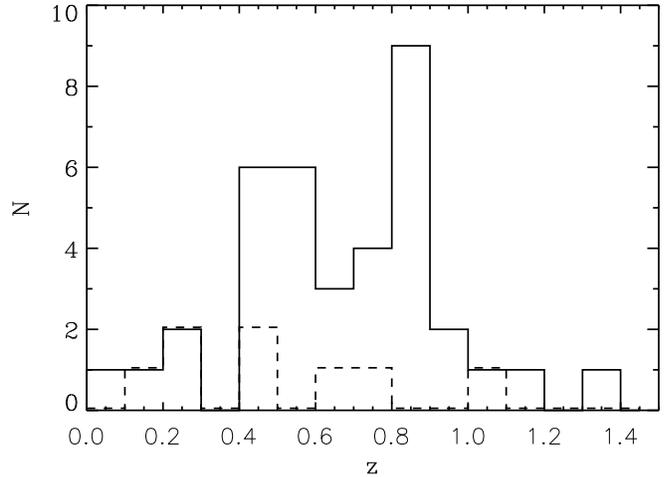}} 
      \caption{Redshift distribution of the additional objects. Labels: 
   solid line - 37 NELGs, dotted line - 8 optically normal galaxies.\label{add_redshift}} 
\end{figure} 

Fig.~\ref{add_redshift} illustrates the redshift distribution of the 
additional non-X-ray emitting objects in the Marano Field. Narrow 
emission line galaxies (NELG) outnumber normal galaxies substantially,
this is due to the fact that a large fraction  of the spectra
without emission lines did not have sufficient SNR to determine a redshift.
Non X-ray emitting NELG peak at $z =0.5-0.9$. 
At $z > 1.0$ the detection of NELG and normal galaxies drops dramatically. 
 
Despite the low number of objects, from Fig.~\ref{add_redshift} 
and \mbox{Fig.~\ref{N_over_z_4panels}~d} we can compare the redshift distributions 
between X-ray emitting type II AGN and non X-ray emitting NELG. 
Both populations show more or less the same distribution up to $z = 1.2$. 
The two groups peak at $z =0.5-0.9$.

The similar redshift distributions could  mean, that 
the X-ray emitting NELGs are drawn from the same population of 
galaxies as the control sample.
On the other hand the similar redshift distribution could also
be due to a large fraction of false matches of NELGs
to our X-ray sources.
A detailed discussion of the number of false matches is given in
Sect.~\ref{falsematch} and shows that false matches make only a minor contribution
to our sample of narrow line AGN.

\section{Discussion\label{section5}}
\subsection{Type I AGN} 

In Sect.~\ref{AGN_discussion} we showed that 
the X-ray selected type I AGN peak at lower redshifts
than the optically selected sample. 
It is interesting to see whether this difference in the redshift 
distributions is due to a redshift dependence of the QSO spectral
energy distributions (SED).

As a parameter which characterizes the SED 
we calculated the optical to X-ray broad band spectral index
$\alpha_{\rm OX}$ between the UV luminosity density at 2500 $\AA$ 
and the X-ray luminosity density at 1 keV (see Sect.~\ref{section3}).

No significant correlation of $\alpha_{\rm OX}$ with redshift
(Fig.~\ref{alpha-z}) or X-ray luminosity can be found.
However, there is a very significant ($P_{\rm NULL}< 10^{-6}$)
correlation of the optical luminosity and $\alpha_{\rm OX}$
(Fig.~\ref{alpha-lopt}).
This correlation has been found in various samples observed with
EINSTEIN (\citealt{avni}), ROSAT (e.g. \citealt{green0},
\citealt{lamer}), and CHANDRA (e.g. \citealt{steffen}).

For a large sample of optically selected AGN, \cite{steffen} computed the
bivariate linear regression coefficients of $\alpha_{\rm OX}$
as a function of $\log(L_{\rm opt})$ and $z$.
They find that $\alpha_{\rm OX}$ is correlated with
$\log(L_{\rm opt})$, but find no significant correlation
with $z$ (see also \citealt{avni} for similar results on earlier
EINSTEIN data).

If  $\alpha_{\rm OX}$ and optical luminosity are correlated, a non-linear
relation between X-ray luminosity and optical luminosity is expected.
Therefore, we plotted the X-ray luminosities versus the optical
luminosity densities (Fig.~\ref{lx-lopt}) and computed the linear
regression coefficients
between $\log(L_{\rm X})$ and $\log(l_{\rm 2500 \AA})$.
Both variables  $\log(L_{\rm X})$ and $\log(l_{\rm 2500 \AA})$ are
measured quantities and neither of them can be regarded as the independent or 
dependent variable.
We used the ordinary least-squares (OLS) bisector algorithm
as described by \citealt{isobe},  which is symmetric regarding the choice 
of independent and dependent variable.

We find a best fit regression  
$\log(L_{\rm X}) = (0.84 \pm 0.08) \times \log(l_{\rm 2500 \AA}) + (19.37 \pm 2.23)$.
This slope is marginally (2$\sigma$ significance) flatter than the $\beta=1.0$ expected for a linear
$L_{\rm X} - l_{\rm opt}$ relation.
With the same method \citealt{steffen} find a slightly flatter slope $\beta = 0.721 \pm 0.011$.

A correlation flatter than $\beta=1$ implies that the optical luminosities in the sample are
spread over a wider range of values than the X-ray luminosities.
This might explain the abovementioned discrepancies of the redshift distributions
of X-ray and optically selected QSOs. The low luminosity objects are more likely
to be detectable in X-rays, while the highest luminosity objects are relatively more
luminous in the optical and therefore detectable at higher redshifts in optical surveys.

\begin{figure}
   \centering 
   \resizebox{\hsize}{!}{ 
   \includegraphics[bbllx=95,bblly=375,bburx=544,bbury=695]{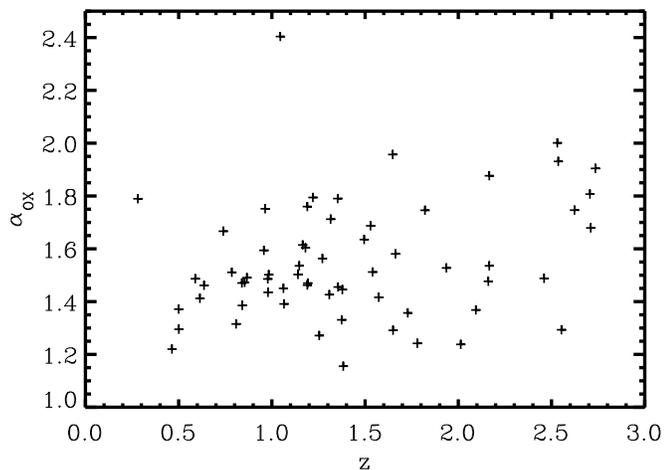}} \caption{UV to X-ray spectral index $\alpha_{\rm OX}$ vs. redshift for 
type I AGN in the core sample.}
         \label{alpha-z} 
\end{figure} 

\begin{figure}
   \centering 
   \resizebox{\hsize}{!}{ 
   \includegraphics[bbllx=95,bblly=375,bburx=544,bbury=695]{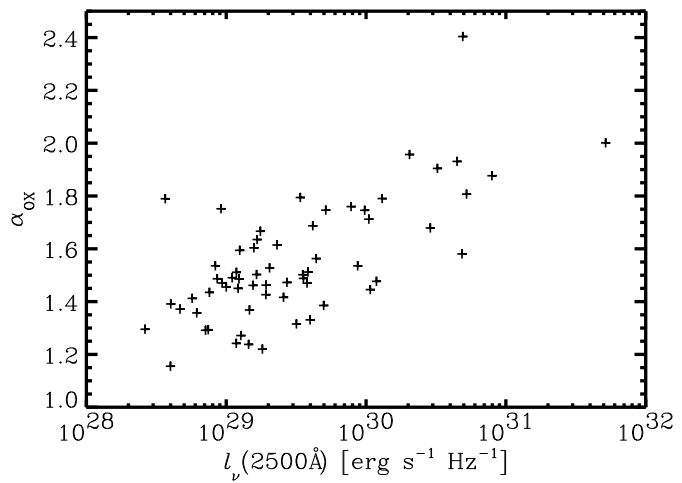}} 
\caption{UV to X-ray spectral index $\alpha_{\rm OX}$ vs. optical luminosity
   density at $2500 \AA$ for type I AGN in the core sample.}
\label{alpha-lopt}

\end{figure}

\begin{figure}
   \centering 
   \resizebox{\hsize}{!}{ 
   \includegraphics[bbllx=95,bblly=375,bburx=544,bbury=695]{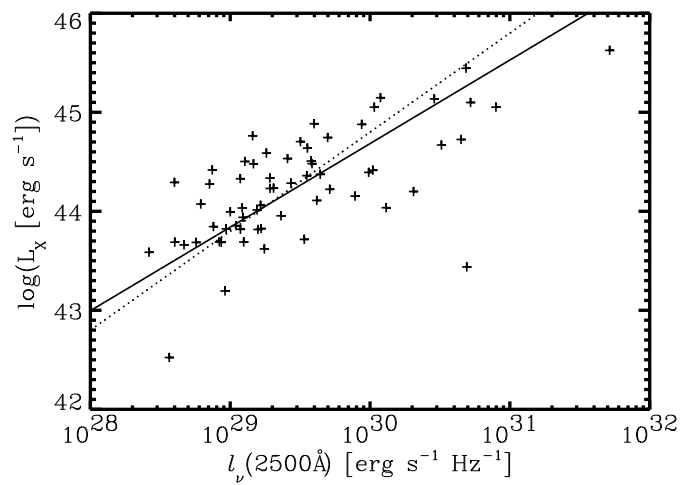}} 
\caption{Observed 0.2-10 keV X-ray luminosity vs. optical luminosity density at 
$2500 \AA$ for type I AGN  in the core sample (crosses). The solid line 
indicates the best fit linear regression with a slope $\beta=0.84$. 
The dotted line with slope $\beta=1$ marks a linear $L_{\rm X} - l_{opt}$ 
relation.}
\label{lx-lopt}
\end{figure}

\subsection{Type II AGN} 
Fig.~\ref{N_over_z_4panels} d shows that most type II AGN are found 
at redshifts
$z\sim 0.5-0.9$, with a few objects at $z> 2.2$. No type II objects were 
identified in the redshift range  $z \sim 1.5 - 2.2$. 
The interesting question is 
whether this redshift gap reflects the intrinsic distribution or is due to 
an observational bias. 

Type II AGN show only narrow emission lines and the optical continuum 
radiation is 
dominated by the host galaxy. The spectroscopic classification
of X-ray sources rely on emission lines like Ly-$\alpha$, 
\ion{C}{iv}, \ion{C}{iii}, \ion{Mg}{ii}, \ion{O}{ii}, and others. Analysis 
of the optical spectra of type II AGN with $z > 0.6$ in our sample, 
in \cite{szokoly}, and in \cite{caccianiga} indicates 
that almost all objects show no \ion{Mg}{ii} emission lines. Only 
in a few cases 
marginal \ion{Mg}{ii} emission is recognized. Effectively, no 
strong spectral features are  present in the wavelength range between
$1909 \AA$ ( \ion{C}{iii}]) and  $3727 \AA$ ([\ion{O}{ii}]).

Since the UV host galaxy continuum is usually very faint, 
already the detection of type II AGN in optical imaging is
hampered by the lack of a strong UV continuum and emission lines, if the 
imaging bandpass falls into this rest frame range.
For $R$-band imaging (5800-7300 $\AA$)  this is the case 
for the redshift range $z \sim 1.0-2.0$.  

Our multi-object spectroscopy covers a useful range of $ \sim 4000 - 9000 \AA$.
This range is reduced, if an object is not centered on the mask,
but offset from the center in the dispersion direction. 
Hence, in many cases only one emission line would be detectable for
type II AGN in the redshift range $\sim 1.0-2.0$. With a low SNR spectrum
this would usually not be sufficient for a clear spectroscopic classification.

For a few sources, which were undetected or very faint in the $R$-band images, we were able to
position a MXU slit  
using the $K$-band counterparts (e.g. 20A, 63A, 463A).
In all of these cases a high redshift type II AGN could be identified.

Fig.~\ref{fig:hr3hr2} supports the assumption that a large fraction
of unidentified sources consists of type II AGN. 
Above \mbox{$HR2=-0.1$} only type II AGN are found as counterparts for identified
X-ray sources. In this region 50\% of the X-ray sources are not 
classified and are expected to be type II AGN.  
Furthermore, Fig.~\ref{fx_R} gives more evidence that the unidentified 
sources overlap with the type II AGN region. 
A ratio \mbox{$f_{\rm{X}}/f_{\rm{opt}}>0.5$} is clearly a sign of AGN activity, 
since normal galaxies and stars usually have $f_{\rm{X}}/f_{\rm{opt}}<0.05$.
Both type II AGN and unidentified objects are found at AGN-typical
$f_{\rm{X}}/f_{\rm{opt}}$, with low X-ray fluxes and faint optical counterparts.

We conclude that the majority of unidentified X-ray sources is likely 
to consist of
type II AGN, most of them presumably with redshifts  $z>1$.
Therefore, the intrinsic redshift distribution of the type II AGN is  uncertain.
The number of unidentified sources is sufficient to fill
the observed gap between $z \sim 1.0 -2.0$.

\subsection{X-ray bright optically normal galaxies}

Five objects in the core sample have been spectroscopically classified 
as galaxies without any emission lines.
The soft X-ray radiation of the low X-ray luminosity galaxies   
(objects 120A, 241A) can be explained by a halo of 
X-ray emitting hot gas around elliptical galaxies (\citealt{sarazin}, \citealt{white}). 

Three optically normal galaxies have X-ray to optical flux ratios and
X-ray luminosities typically found for AGN. Objects of this type have 
been named X-ray bright optically normal galaxies (XBONGs; \citealt{comastri}).  
The objects 8A, 49A, and 204A  have  intermediate redshifts ($0.3<z<0.9$)
and fairly high X-ray luminosities (log$(L_{\rm{X}}) > 43$), rather hard 
X-ray spectra, and  faint optical counterparts ($R>19$).
The high X-ray luminosities and $f_{\rm{X}}/f_{\rm{opt}} \sim 1$ indicate active nuclei. 
The computed 
hydrogen column densities show moderate absorption. 

The properties of our XBONGs are in agreement with other 
studies  (\citealt{silverman}, \citealt{severgnini}).
Even though different scenarios are discussed, the nature of XBONGs remains 
still unclear (\citealt{brandt}).
\cite{comastri} assume that the non-detection of optical emission
lines and the hard X-ray colors are due to a heavily absorbed AGN embedded 
in a galaxy whose X-ray emission is due to a scattered/reprocessed 
nuclear component.
The three XBONGs do not show harder $HR2$ values than type II AGN 
and also do not have the highest column densities of the AGN sample. Hence,
a heavily absorbed AGN without scattered/reprocessed X-ray radiation is
ruled out. Scattered or reprocessed radiation is necessary to explain the
X-ray and optical observations. Following \cite{komossa}, the observed 
X-ray luminosity of a scattered emission by a warm reflector is only a 
hundredth or thousandth of the intrinsic X-ray luminosity. For our objects 
with observed \mbox{log$(L_{\rm{X}}) \sim 43-44$,} this would require intrinsic X-ray 
luminosities of \mbox{log$(L_{\rm{X}}) \sim 45-47$}, which would make them by far the 
most luminous X-ray emitters in the sample. 

Another possible explanation for these objects is given by \cite{severgnini}. 
They studied three low redshift XBONGs with log$(L_{\rm{X}}) \sim 42-43$ and 
find $N_{\rm H}$ values similar to our sample. In their interpretation 
the faint emission lines of an obscured or unobscured AGN up to an intrinsic log$(L_{\rm{X}}) \sim 43$ 
can be overwhelmed by a host galaxy with an absolute magnitude $M_{\rm R}\geq -22$.
However, XBONGs 8A and 49A are not consistent with the scenario mentioned by 
\cite{severgnini}. 
The AGN of 49A with an intrinsic log$(L_{\rm{X}}) = 43.7$ should be 
optically too bright to be hidden by a galaxy of $M_{\rm R}=-21.8$. 
Moreover, source 8A, which is also detected as a radio source (\citealt{gruppioni}),
has an intrinsic log$(L_{\rm{X}}) = 43.9$ 
and is almost one magnitude X-ray brighter than the examples in \cite{severgnini}. 
The absolute magnitude of the host galaxy is $M_{\rm R}=-22.6$, 
but an optically much brighter galaxy is needed to 
hide the emission lines of such a powerful AGN.
By adding a template type I spectrum to the measured spectra
of objects 8A and 49A we estimated that 
any hidden type I AGN in these objects would have 
$\alpha_{\rm OX}$ values of $\sim$0.8 or less.

Regarding the X-ray and optical colors, our  XBONGs
are very similar to type II AGN. Therefore, it is likely that these 
sources have a narrow line type II spectrum, which is intrinsically
weak or dust absorbed and  not detected above the continuum of the
host galaxy. 

This result is consistent with \cite{caccianiga} who state that  
more accurate re-observation (high resolution data and/or better spectral 
coverage) of hard X-ray emitting galaxies will reveal narrow emission 
lines and, therefore, their real AGN nature. \cite{severgnini} also claim 
possible misclassification of type II AGN as XBONGs. However, some of
these objects are still classified as XBONGs after a high resolution
observation with better spectral coverage.

As for type II AGN, the observed redshift distribution of XBONGs could be due to
an observational bias. XBONGs are found with brighter R-magnitudes than
typical type II
AGN up to $z=1$. But missing emission line features make them optically even 
more difficult to identify than type II AGN. For a reliable classification
they have to be optically brighter, thus limiting their maximum redshift.

\subsection{Stars}
In the Marano Field survey 7\% of the X-ray sources are identified with galactic stars.
As in other deep surveys, these are typically G, K, and M stars, 
whose X-ray emission is caused by magnetic activity (\citealt{brandt}). 
Two sources classified as stars have  $f_{\rm{X}}/f_{\rm{opt}} \sim 1$ (Fig.~\ref{fx_R}),
which is more typical for AGN.
However, we carefully reanalyzed the
X-ray colors, optical properties, and the optical image. 
Apart from the unusally high $f_{\rm{X}}/f_{\rm{opt}}$ all existing evidence suggests
M-stars as reliable counterparts for both X-ray sources.
 
\section{Conclusions \label{section7}} 

With a total of 120 ksec good observation time we detect 328 X-ray
sources. Among 140 spectroscopic classifications of 187 optical
counterparts (in a 3$\sigma_{\rm{X}}$ position
error with $\sigma_{\rm{syst.}} =0.7\arcsec$) to 328 X-ray sources (not
completely covered by optical data), we find
89 broad emission line objects, 36 narrow emission line objects, 6 galaxies, 
and 9 stars. In the central region of the Marano Field we reach an 
identification completeness of 65\%. 

While the redshift distribution of the optically selected QSOs in the field 
is basically flat up to $z\simeq 3$, the distribution of the 
XMM-Newton sources peaks at  $z\simeq 1$.
Using our sample of XMM-Newton sources classified as type I AGN, 
we investigate possible causes for this tendency of 
deep X-ray surveys to discover faint populations at 
comparably low redshifts. 
We find no significant correlation of the optical to 
X-ray SED slope $\alpha_{\rm OX}$ with redshift. 
As it is widely reported in the literature, $\alpha_{\rm OX}$ is tightly correlated 
with optical luminosity. 
A different representation of this correlation is 
the non-linear dependency $L_{\rm X}(l_{\rm opt})$.
The best fit regression  $\log(L_{\rm X}) = 0.84 \times \log(l_{\rm 2500 \AA}) + 19.36$
implies that the optical luminosities in a typical sample spread over a wider
range than the X-ray luminosities.
Therefore, the less luminous objects of the population are more easily detected in 
X-rays than in the optical. 
On the other hand, an increase of optical luminosity is, on average, not
accompanied by a proportional increase in X-ray luminosity.
Hence, the luminous (and more distant) objects are detected  more efficiently
in optical surveys. 

In the core region of the field we classified 31 new type II AGN.
Most of them are found at redshifts $z<1.5$;
additionally we find five high redshift type II
AGN at $z>2.2$. Fifteen objects can be classified as type II QSOs
with intrinsic X-ray luminosities $L_{\rm X} > 10^{44} {\rm erg/s}$. 

We show that the optical identification of type II AGN
is very difficult in the redshift range  $z=1-2$ due to the absence 
of suitable emission lines
in the optical window. Therefore, their intrinsic redshift distribution 
remains unclear. We demonstrate that the use of $K$-band data for 
MXU slit positioning reveals type II AGN or XBONGs that 
would have likely been missed in $R$-band images. 

The X-ray selected type II AGN have a very similar redshift
distribution as non X-ray emitting narrow emission line galaxies,
which have been spectroscopically classified in the same field as a control sample.

The intrinsic hydrogen column densities, as derived from X-ray hardness ratios,
show that the fraction of 
absorbed X-ray sources is much higher for type II AGN than for type I. 
Nevertheless, we find a few unabsorbed type II AGN and some evidence for
absorption in high redshift type I AGN.
However, due to the faintness of the sources,
the significance of absorption in the individual type I AGN is low.
Furthermore,  at high redshifts statistical fluctuations in the
X-ray spectrum can lead to high values of spuriously measured
$N_{\rm H}$  values (e.g. \citealt{akylas}). 
If we only include $2\sigma$ detections of intrinsic $N_{\rm H}$ 
in our analysis, only 4 absorbed type I AGN remain and no 
dependency of absorbed fraction on redshift is obvious.
In the CHANDRA data Chandra Deep Field South (CDFS) 
 \citealt{tozzi} find hints of an increase of absorbed fraction with 
redshift. However, these data probably also suffer from the uncertainties 
mentioned above. Using XMM data in the same field, \cite{dwelly} find 
litte evidence that the absorption distribution is dependent on
either intrinsic X-ray luminosity or redshift.

Our type I and type II AGN cover the same range in absorption  corrected 
X-ray luminosity. However, the mean corrected X-ray 
luminosity is smaller for type II AGN than 
for type I AGN. 

Three of the XMM-Newton classifications are X-ray bright optically 
normal galaxies (XBONGs), which show X-ray 
luminosities typical for AGN, but no optical emission lines. 
Their X-ray luminosities of log$(L_{\rm{X}}) \sim 43-44$ are comparable to 
the mean type II AGN X-ray luminosity. They do not show harder X-ray 
spectra and do not reveal higher hydrogen column densities than the average type II AGN.
We conclude that the objects are very similar to type II AGN. 
However, their narrow emission lines are not detected, since they are either
intrinsically weak or obscured by dust.

\begin{acknowledgements} 
Mirko Krumpe is supported by the Deutsches  
Zentrum f\"ur Luft- und Raumfahrt (DLR) GmbH  
under contract No. FKZ 50 OR 0404. 
Georg Lamer acknowledges support by the Deutsches Zentrum f\"ur Luft- und  
Raumfahrt (DLR) GmbH under contract no.~FKZ 50 OX 0201. 
\end{acknowledgements}

\begin{landscape}
 \setlength{\headsep}{160pt} 
               \begin{longtable}{rrlcccccccccccc} 
\caption{Optical properties of candidate counterparts of Marano XMM-Newton 
  X-ray sources. For a detailed explanation of the individual columns see 
  Sect.~\ref{section3}. \label{opt_tab}} \\ \hline  
   (1)&(2)&(3)&(4)&(5)&(6)&(7)&(8)&(9)&(10)&(11)&(12)&(13)&(14)&(15)\\  
   $No$ &$RA$&$DEC$&$dist_{\rm{OX}}$&$K$&$R$&$class$&$z$&$flags$&$log(L_{\rm{X_{OBS}}}$)&$M_{\rm{B}}$&$\alpha_{\rm{OX}}$&$N_{\rm H}$&$log(L_{\rm{X_{INT}}}$)&$rem.$\\  
   \hline
\endfirsthead
\caption{continued}\\\hline
    (1)&(2)&(3)&(4)&(5)&(6)&(7)&(8)&(9)&(10)&(11)&(12)&(13)&(14)&(15)\\  
   $No$&$RA$&$DEC$&$dist_{\rm{OX}}$&$K$&$R$&$class$&$z$&$flags$&$log(L_{\rm{X_{OBS}}}$)&$M_{\rm{B}}$&$\alpha_{\rm{OX}}$&$N_{\rm
   H}$&$log(L_{\rm{X_{INT}}}$)&$rem.$\\ 
\hline
\endhead
\hline
\endfoot
  1A& 3 15 49.6 & -55 18 12& 1.52& 17.14&  20.39&  B&0.808 &1\,-\,-&44.70&-22.26& 1.32&0.00$_{-0.00}^{+0.01}$ & 44.70&        2,3\\ 
  2A& 3 15 47.5 & -55 29 04& 1.37& --&  19.66&  B&0.464 &1\,-\,- &44.59&-21.66& 1.22&0.00$_{-0.00}^{+0.04}$ & 44.59&               2\\ 
  3A& 3 17 32.7 & -55 20 26& 1.35& --&  --&  B&0.406 &1\,-\,-&44.24& --&--&0.28$_{-0.12}^{+0.10}$ & 44.44&              2\\ 
  4A& 3 13 34.0 & -55 26 43& 1.26& --&  17.05&  B&0.987&111 &45.28&-26.04& 1.74&0.00$_{-0.00}^{+0.01}$ & 45.28&               1\\ 
  5A& 3 16 50.4 & -55 11 09& 0.78& --&  17.51&  B&2.531&1\,-\,- &45.63&-27.79& 2.00&0.00$_{-0.00}^{+0.01}$ & 45.63&       1,3\\ 
  6A& 3 16 05.8 & -55 15 39& 1.12& 17.79&  20.64&  B&0.636 &111&44.01&-21.49& 1.46&0.00$_{-0.00}^{+0.01}$ & 44.01&       1,3\\ 
  7A& 3 15 05.6 & -55 09 42& 0.13& 18.25&  21.98&  B&0.501 &111&43.59&-19.56& 1.30&0.00$_{-0.00}^{+0.01}$ & 43.59&              4\\ 
  8A& 3 14 56.0 & -55 20 07& 1.17& 15.53&  19.03&  G&0.387 &111&43.32& --&--&2.09$_{-0.73}^{+0.47}$ & 43.90&       3,5\\ 
  9A& 3 15 09.9 & -55 13 13& 1.18& 18.81&  22.63&  N&1.427&111    &44.64& --&--& 0.33$_{-0.20}^{+0.18}$ & 44.73&             4,C\\ 
 10A& 3 13 28.3 & -55 10 18& 0.49& 16.98&  20.24&  B&1.378 &111&45.05&-23.58& 1.45&0.23$_{-0.16}^{+0.18}$ & 45.12&               3\\ 
 11A& 3 15 11.3 & -55 09 27& 0.35& 17.98&  21.02&  B&1.192 &111&44.51&-22.45& 1.47&0.00$_{-0.00}^{+0.01}$ & 44.51&       2,3\\ 
 12A& 3 14 32.4 & -55 14 40& 0.46&  8.87&  --&  S&0.000 &1\,-\,-&  --& --&--& 0.00$_{-0.00}^{+0.01}$ & --&          3\\ 
 13A& 3 13 50.9 & -55 18 38& 1.54& 18.78&  21.35&  B&0.500 &1\,-\,-&43.66&-20.18& 1.37&0.00$_{-0.00}^{+0.01}$&       43.66&               3\\ 
 14A& 3 16 38.0 & -55 06 37& 1.71& 17.47&  20.68&  B&0.854 &1\,-\,-&44.28&-22.09& 1.47& 0.00$_{-0.00}^{+0.01}$&       44.28&       2,3\\ 
 15A& 3 15 51.7 & -55 08 17& 9.58& --&  21.69&        N &      0.331    &000&--& --&--&   --&--&             4, C\\ 
 16A& 3 15 38.3 & -55 01 40& 0.41& 19.40&  21.30&  B&1.374 &111&44.88&-22.51& 1.33&0.00$_{-0.00}^{+0.01}$&       44.88&               3\\ 
 17A& 3 15 28.9 & -55 10 27& 3.62& 18.49&  21.27&  B&2.161&111&45.15&-23.69& 1.48& 0.00$_{-0.00}^{+0.19}$&       45.15&2,3,C\\ 
 18A& 3 14 21.1 & -55 24 05& 2.02& --&  21.65&  B&0.614 &1\,-\,-&43.69&-20.40& 1.41&0.05$_{-0.05}^{+0.08}$&       43.72&               3\\ 
 19A& 3 15 25.2 & -55 18 27& 0.82& 19.78&  22.13&  B&1.573 &111&44.53&-22.03& 1.42& 0.00$_{-0.00}^{+0.16}$&       44.53&              3\\ 
 20A& 3 16 21.5 & -55 17 59& 0.63& 17.33&  --&  N&2.207&111&45.01& --&--&0.83$_{-0.44}^{+0.52}$&       45.12&              C\\ 
 20B& 3 16 21.1 & -55 18 01& 4.11& 13.28&  16.72&        S &      0.000&011  &--& --&--&  --&--&              3\\ 
 21A& 3 16 26.1 & -55 22 51& 1.38& 12.16&  16.50&  S&0.000&1\,-\,-    &--& --&--&  0.00$_{-0.00}^{+0.02}$ & --&             3\\ 
 22A& 3 15 19.7 & -55 02 25& 4.63& 19.41&  23.33&  B&1.78 &111&44.63&-21.19& 1.11&0.00$_{-0.00}^{+0.10}$&       44.63&              C\\ 
 22B& 3 15 20.1 & -55 02 34& 4.62& 17.36&  19.98&  B&1.353 &111&44.34&-23.79& 1.66&0.00$_{-0.00}^{+0.06}$&       44.34&1,3,C\\ 
 23A& 3 14 32.1 & -55 19 59& 1.24& 18.32&  21.01&  B&1.271 &111&44.37&-22.62& 1.56&0.00$_{-0.00}^{+0.02}$&       44.37&               1\\ 
 25A& 3 15 34.7 & -55 19 26& 0.85& 18.12&  --&  B&1.430 &111&44.48& --&--&0.00$_{-0.00}^{+0.03}$&       44.48&               3\\ 
 26A& 3 14 36.0 & -55 14 03& 0.80& 17.90&  20.23&  B&1.19&111   &44.15&-23.24& 1.76&  0.00$_{-0.00}^{+0.03}$&       44.15&             4,C\\ 
 27A& 3 15 38.7 & -55 22 19& 1.74& 18.04&  21.75&  B&1.190 &111&44.23&-21.72& 1.46& 0.00$_{-0.00}^{+0.01}$&       44.23&              3\\ 
 28A& 3 15 07.6 & -55 04 56& 0.64& 17.75&  20.15&  B&1.315 &111&44.41&-23.56& 1.71&0.00$_{-0.00}^{+0.01}$&       44.41&        2,3\\ 
 29A& 3 15 11.1 & -55 15 30& 0.42&  9.28&  --&  S&0.000 &1\,-\,-&--& --&--&  0.00$_{-0.00}^{+0.01}$&  --&            3\\ 
 32A& 3 15 47.0 & -55 17 55& 1.44& 17.75&  22.96&  N&2.727 &110   &45.09& --&--&  14.5$_{-8.2}^{+7.9}$&       45.55&            C      \\ 
 33A& 3 13 19.4 & -55 16 17& 0.62& 17.18&  19.21&  B&2.166 &1\,-\,-&45.05&-25.76& 1.88& 0.10$_{-0.10}^{+0.33}$ & 45.07&             1\\ 
 34A& 3 15 59.1 & -55 26 37& 1.72& 18.32&  20.70&  B&0.983 &1\,-\,-&44.36&-22.38& 1.50& 0.00$_{-0.00}^{+0.01}$ & 44.36&             3\\ 
 35A& 3 15 03.2 & -55 19 06& 0.67& 16.17&  19.54&  N&0.391 &111&42.91& --&--& 1.32$_{-0.58}^{+0.59}$ & 43.39&      4, C\\ 
 35B& 3 15 03.2 & -55 19 09& 3.28& --&  19.54&        N &      0.391&011  &--& --&--& --&--&             C\\ 
 36A& 3 14 11.1 & -55 18 29& 3.12& 19.91&  --&       -- &-- &     --  &--& --&--&  --&--&             4\\ 
 38A& 3 16 30.0 & -55 19 11& 2.28& 19.66&  22.33&  B&1.254 &111&44.50&-21.27& 1.27& 0.00$_{-0.00}^{+0.01}$&       44.50&       2,3\\ 
 39A& 3 13 39.7 & -55 01 51& 0.95& 19.24&  23.41&  N&0.862 &110&44.07& --&--&0.76$_{-0.33}^{+0.63}$ & 44.32&                   \\ 
 41A& 3 16 07.9 & -55 17 24& 1.58& 18.54&  22.36&  B&0.979 &111&43.85&-20.71& 1.43& 0.00$_{-0.00}^{+0.05}$ & 43.85&              4\\ 
 42A& 3 15 48.7 & -55 22 46& 0.94& 19.37&  21.63&  B&1.54&111    &44.48&-22.47& 1.51& 0.00$_{-0.00}^{+0.02}$&       44.48&              4,C\\ 
 43A& 3 13 32.5 & -55 10 49& 1.14& 19.52&  23.40&  B&2.013 &111&44.76&-21.41& 1.24&0.17$_{-0.17}^{+0.32}$ & 44.80&                   \\ 
 44A& 3 13 46.8 & -55 00 31& 1.07& 17.42&  21.40&  B&0.785 &111&43.82&-21.19& 1.51& 0.00$_{-0.00}^{+0.10}$&       43.82&                   \\ 
 44B& 3 13 47.4 & -55 00 30& 6.41& --&  22.24&        N &      0.480&000 &--& --&--&  --&--&                  \\ 
 45A& 3 14 51.8 & -55 19 32& 2.49& --&  23.09&  N&  0.584&100&43.08& --&--&0.69$_{-0.27}^{+0.51}$ & 43.38&              4\\ 
 46A& 3 16 07.0 & -55 23 24& 4.01& 19.16&  21.98&  B&1.308&111&44.34&-21.71& 1.43&0.00$_{-0.00}^{+0.09}$ & 44.34&            C\\ 
 47A& 3 15 38.7 & -55 10 44& 1.34& 19.06&  23.94&  N&0.900&100&43.47& --&--& 2.09$_{-1.09}^{+1.54}$ & 43.89&             C\\ 
 47B& 3 15 39.1 & -55 10 42& 3.72& 19.99&  22.76&        N &      0.89&000 &--& --&--&  --&--&            C\\ 
 49A& 3 16 38.3 & -55 20 14& 1.66& 16.01&  20.26&  G&0.454 &111&43.22& --&--&1.45$_{-0.63}^{+0.65}$ & 43.70&                  \\ 
 50A& 3 14 09.9 & -55 17 46& 1.25& 18.35&  23.66&  N&0.986 &110&43.79& --&--&3.63$_{-1.58}^{+1.63}$ & 44.30&                  \\ 
 51A& 3 16 30.6 & -55 15 03& 1.83& 16.57&  20.95&  N&0.58 &110&43.37&--& --&0.05$_{-0.05}^{+0.08}$ & 43.41&                   \\ 
 52A& 3 14 51.4 & -54 57 14& 1.82& --&  --&  B&1.829 &1\,-\,-&44.77& --&--&0.00$_{-0.00}^{+0.04}$ & 44.77&              2\\ 
 53A& 3 13 58.8 & -55 17 54& 0.67& 18.82&  23.23&  N&1.340&100&44.07&--& --&3.98$_{-2.93}^{+2.51}$ & 44.50&                   \\ 
 54A& 3 16 12.4 & -54 59 10& 2.49& --&  --&  B&2.744 &1\,-\,-&45.05&    --&--&0.00$_{-0.00}^{+0.40}$ & 45.05&            1\\ 
 56A& 3 12 50.9 & -55 17 26& 1.94& 16.65&  20.11&  B&0.641 &111&43.90&-22.04& 1.60&0.01$_{-0.01}^{+0.08}$ & 43.91&              1\\ 
 57A& 3 13 44.0 & -55 19 24& 2.33& 17.11&  20.12&  B&0.281 &1\,-\,-   &42.52&-19.93& 1.79&  0.01$_{-0.01}^{+0.08}$&       42.54&             3\\ 
 58A& 3 13 01.7 & -55 22 23& 0.70& --&  21.10&  B&0.589&111    &43.69&-20.85& 1.49&0.00$_{-0.00}^{+0.02}$ & 43.69& \\
 59A& 3 13 39.4 & -55 14 26& 0.77& 17.37&  21.83&  B&0.978&111    &43.94&-21.23& 1.49&0.00$_{-0.00}^{+0.02}$ & 43.94&\\
 60A& 3 14 26.3 & -55 17 47& 0.71& 17.62&  21.77&  B&0.957 &111&43.69&-21.25& 1.59&0.03$_{-0.03}^{+0.12}$ & 43.71&3,5\\ 
 63A& 3 15 16.9 & -55 06 02& 1.25& 20.37&  23.19&  N&2.800 &110&44.88& --&--&4.37$_{-2.10}^{+11.2}$ & 45.13&               C   \\ 
 64A& 3 13 33.7 & -55 10 21& 2.32& 17.21&  21.51&  B&1.165 &111&43.95&-21.91& 1.61&0.00$_{-0.00}^{+0.08}$ & 43.95&                   \\ 
 66A& 3 15 00.7 & -55 07 18& 1.23& 18.40&  21.41&  N&0.981&100&43.51& --&--& 0.28$_{-0.20}^{+0.28}$ & 43.63&  C\\ 
 67A& 3 13 10.6 & -55 13 13& 1.22& 18.27&  21.69&  B&0.866 &111&43.85&-21.11& 1.49&0.00$_{-0.00}^{+0.03}$ & 43.85&                    \\ 
 68A& 3 13 37.9 & -55 23 03& 1.77& 18.28&  21.80&  B&0.839&111&43.82&-20.93& 1.47&0.00$_{-0.00}^{+0.08}$ & 43.82&              5\\ 
 69A& 3 14 15.3 & -55 19 10& 0.50& 19.37&  24.63&  B&2.553&111   &44.42&-20.69& 1.29&1.20$_{-0.77}^{+1.21}$ & 44.53&                   \\ 
 70A& 3 16 21.0 & -55 06 49& 1.98& 19.82&  23.37&       -- &  -- &    --  &--& --&--&  --&--&                   \\ 
 70B& 3 16 21.0 & -55 06 45& 3.35& 18.96&  22.15&        N &      0.844&010  &--& --&--&  --&--&                   \\ 
 73A& 3 13 21.2 & -55 20 47& 0.51& 18.31&  20.10&  B&2.706 &111  &45.10&-25.31& 1.81& 0.00$_{-0.00}^{+0.12}$ & 45.10&              1\\ 
 75A& 3 16 01.2 & -55 05 16& 0.50& 18.96&  22.86&  B&2.460&111  &44.64&-22.38& 1.49&0.00$_{-0.00}^{+0.34}$ & 44.64&                    \\ 
 76A& 3 15 49.7 & -55 09 08& 0.16& 19.06&  23.21&  B&1.065 &111&43.69&-20.01& 1.39&0.01$_{-0.01}^{+0.15}$ & 43.69&               3\\ 
 78A& 3 15 02.0 & -55 26 11& 1.13& 19.17&  23.68&       -- &     --  & --  &--& --&--&  --&--&                  \\ 
 80A& 3 13 59.4 & -54 57 17& 3.16& --&  --&  B&1.620&1\,-\,-&44.51& --&--&0.00$_{-0.00}^{+0.05}$ & 44.51&               1\\ 
 81A& 3 13 44.2 & -55 01 52& 4.16& 18.86&  --&       -- &     --  &--& --&--&  --&--&                  \\ 
 83A& 3 14 47.9 & -55 22 31& 2.33& 19.46&  23.68&  B&1.650 &111&44.27&-20.63& 1.29&0.00$_{-0.00}^{+0.07}$&44.27&                    \\ 
 83B& 3 14 48.8 & -55 22 30& 5.51& 18.40&  20.70&        N &      0.568  &010&--& --&--& --&--&              3\\ 
 84A& 3 16 21.6 & -55 20 38& 1.32& 18.63&  23.47&  B&2.094&111   &44.48&-21.43& 1.37&1.32$_{-0.75}^{+1.27}$ & 44.64&                   \\ 
 84B& 3 16 21.9 & -55 20 41& 3.72& 18.73&  23.42&        S &      0.000  &011&--& --&--&  --&--&                   \\ 
 88A& 3 13 16.6 & -55 03 10& 3.31& --&  20.85&  B&0.739 &111&43.62&-21.62& 1.67&0.00$_{-0.00}^{+0.03}$ & 43.62&                    \\ 
 92A& 3 12 00.3 & -55 02 22& 1.85& --&  --&  B&2.425 &1\,-\,-&45.22& --&--&0.58$_{-0.58}^{+0.97}$ & 45.29&              1\\ 
100A& 3 12 43.8 & -55 10 58& 2.43& --&  --&  B&1.165 &111&44.23&    --&--& 0.01$_{-0.01}^{+0.17}$ & 44.24&                   \\ 
104A& 3 16 47.0 & -55 10 33& 3.80& --&  23.86&       -- &      --  &--  &--& --&--&--&--&                    \\ 
111A& 3 15 05.2 & -55 29 48& 2.42& --&  --&       -- &     --  & --  &--& --&--&  --&--&                  \\ 
116A& 3 16 20.7 & -55 16 52& 1.30& 16.99&  --&  N&0.581 &111&43.08& --&--&3.63$_{-1.42}^{+1.37}$ & 43.72&                   \\ 
120A& 3 13 13.5 & -55 01 59& 0.87& --&  15.90&  G&0.094 &1\,-\,-&41.46& --&--&0.00$_{-0.00}^{+0.03}$ & 41.46&                   \\ 
132A& 3 16 14.3 & -55 17 29& 1.53& 17.65&  21.79&  N&1.144&100&43.70& --&--&0.00$_{-0.00}^{+0.18}$ & 43.70&  \\ 
133A& 3 14 26.2 & -55 21 13& 1.02& --&  23.54&  N&2.321 &110&44.25& --&--&0.00$_{-0.00}^{+0.20}$ & 44.25&            C       \\ 
135A& 3 13 14.7 & -55 26 19& 5.41& --&  21.95&  B&2.033 &111&44.79&-22.88& 1.48&0.00$_{-0.00}^{+0.10}$ & 44.79&                    \\ 
140A& 3 14 33.0 & -55 25 18& 0.91& --&  21.84&       -- &     --  &--  &--& --&--& --&--&                    \\ 
145A& 3 15 37.1 & -55 17 14& 1.40& 16.39&  19.77&  N&0.497 &111&42.31& --&--&100$_{-31}$ & 44.11&                   \\ 
150A& 3 14 29.5 & -55 06 05& 0.32& 18.15&  23.21&       -- & -- & --  &--& --&--&   --&--&                  \\ 
151A& 3 16 50.5 & -55 11 01& 4.45& --&  --&       -- &     --  &--& --&--&  --&--& --&            C\\ 
157A& 3 16 28.0 & -55 05 38& 3.46& 16.41&  19.76&  N&0.625 &111&42.79& --&--&--&--&                   \\ 
165A& 3 15 32.2 & -55 16 54& 5.06& 10.94&  12.90&        S &      0.000&011  &--& --&--&--&--&                    \\ 
166A& 3 14 55.3 & -55 18 15& 6.85& 17.91&  21.45&        S &      0.000  &011&--& --&--&--&--&                    \\ 
171A& 3 13 51.4 & -55 02 56& 2.18& 18.59&  22.17&  N&0.800 &111   &43.45& --&--&6.31$_{-2.47}^{+2.38}$ & 44.14&                    \\ 
172A& 3 16 24.3 & -55 19 08& 3.17& --&  --&       -- &     --  &--&--& --&--&--&--&                    \\ 
185A& 3 16 50.8 & -55 22 24& 1.60& --&  --&  B&2.768 &1\,-\,-&44.90&    --&--&6.31$_{-4.03}^{+7.80}$ & 45.21&               2\\ 
189A& 3 13 25.4 & -55 01 18& 8.94& 19.34&  20.47&        N &0.079&011 &--& --&--&--&--&                    \\ 
191A& 3 14 38.5 & -55 20 06& 2.27& 15.98&  17.95&  B&1.045&1\,-\,-&43.44&-25.24& 2.40&5.75$_{-2.81}^{+3.83}$ & 44.01&              1,C\\ 
197A& 3 14 27.9 & -55 19 34& 2.27& 17.94&  --&       -- &--  &--  &--& --&--& --&--&                   \\ 
204A& 3 14 24.6 & -55 02 03& 1.31& 18.14&  21.16&  G&0.859&110    &43.30& --&--&0.03$_{-0.03}^{+2.26}$ & 43.32&              \\ 
205A& 3 13 18.9 & -55 25 12& 1.73& --&  --&  S&0.000&111&--&   --&--&0.00$_{-0.00}^{+0.01}$ & --&                    \\ 
205B& 3 13 18.8 & -55 25 07& 4.20& --&  --&        S &      0.000  &011&--& --&--& --&--&                    \\ 
209A& 3 16 02.2 & -55 02 48& 0.88& 19.31&  20.63&  B&2.735&111    &44.67&-24.79& 1.90&1.58$_{-1.17}^{+2.50}$ & 44.80&               2\\ 
217A& 3 14 19.6 & -55 16 43& 3.81& 19.35&  22.07&  N&0.816&110&42.96& --&--&3.63$_{-1.77}^{+2.42}$ & 43.51&              C\\ 
217B& 3 14 19.2 & -55 16 37& 3.66& 18.00&  21.32&        N &      0.645&010  &--& --&--&--&--&              C\\ 
220A& 3 15 50.4 & -55 01 45& 1.75& 18.48&  21.53&  B&1.529 &111&44.11&-22.55& 1.69&0.00$_{-0.00}^{+0.10}$ & 44.11&                    \\ 
222A& 3 16 43.2 & -55 20 06& 8.05& --&  --&        G &      0.82&000&--& --&--& --&--&                   \\ 
223A& 3 16 47.6 & -55 14 11& 1.37& --&  --&  N&1.304&100&44.07& --&--&0.28$_{-0.25}^{+0.45}$ & 44.16&                    \\ 
224A& 3 13 04.9 & -55 16 07& 1.89& 18.36&  21.46&        G &      0.688&001    &--& --&--& --&--&         C          \\ 
224B& 3 13 04.8 & -55 16 04& 1.70& 17.48&  21.51&  N&0.690&111    &43.48& --&--&17.4$_{-5.1}^{+7.9}$ & 44.48&            C      \\ 
225A& 3 13 31.7 & -55 00 46& 3.12& 15.89&  --&        G &      0.420&011   &--& --&--& --&--&     \\
225B& 3 13 32.2 & -55 00 45& 1.32& 16.06&  17.99&        S &      0.000 &011 &--& --&--&--&--&                    \\ 
229A& 3 16 06.5 & -55 14 44& 4.07& --&  22.23&  N&0.98&001&--& --&--& --&--&                  \\
232A& 3 14 41.7 & -55 08 17& 1.74& 19.48&  --&  B&2.520&111&44.58& --&--&1.20$_{-0.95}^{+16.4}$ & 44.71&                   \\ 
241A& 3 13 21.9 & -55 13 51& 0.95& 15.81&  17.56&  G&0.165&111&41.85& --&--&0.01$_{-0.01}^{+0.06}$ & 41.87&                    \\ 
242A& 3 16 04.2 & -55 07 16& 0.81& 18.34&  22.57&  B&1.147 &111&43.69&-20.81& 1.54& 0.40$_{-0.25}^{+0.47}$ & 43.82&                   \\ 
246A& 3 13 56.6 & -55 01 07& 2.96& 18.50&  21.91&  N&0.944 &110   &43.51& --&--&0.52$_{-0.35}^{+1.77}$ & 43.70&                    \\ 
246B& 3 13 57.1 & -55 01 13& 4.75& 20.25&  22.64&        N &      0.688&0\,-\,-  &--& --&--&  --&--&    6\\ 
253A& 3 14 38.0 & -55 06 50& 2.64& 18.21&  21.84&  N&0.517 &110&42.64& --&--&25.1$_{-15.2}^{+28.5}$ & 43.83&              C\\ 
253B& 3 14 38.1 & -55 06 45& 3.12& --&  23.86&        N &      0.518&010  &--& --&--&--&--&              C\\ 
255A& 3 14 46.2 & -55 09 52& 4.28& 14.66&  17.51&        S & 0.00&011  &--& --&--& --&--&                   \\ 
265A& 3 13 36.7 & -55 00 18& 0.94& 18.89&  22.65&  B&1.280&111&43.74&-21.00& 1.56&0.91$_{-0.90}^{+3.29}$ & 43.94&                  \\ 
265B& 3 13 37.0 & -55 00 20& 4.04& 19.84&  21.81&        S &      0.000&011  &--& --&--& --&--&                   \\ 
267A& 3 14 22.6 & -55 17 09& 2.71& 18.53&  22.46&  B&1.495&111&43.83&-21.56& 1.64&8.32$_{-4.06}^{+6.68}$ & 44.37&                    \\ 
268A& 3 16 23.8 & -55 15 20& 1.17& 18.06&  21.20&  B&1.22&111&43.72&-22.34& 1.79&0.00$_{-0.00}^{+0.10}$ & 43.72&               4\\ 
268B& 3 16 23.9 & -55 15 21& 1.68& 17.82&  21.20&        N &      0.983&010  &--& --&--& --&--&                   \\ 
280A& 3 13 24.8 & -55 11 19& 0.66& 18.93&  22.95&  B&1.936 &110&44.23&-21.78& 1.53&27.5$_{-12.6}^{+21.1}$ & 44.97&         C           \\ 
281A& 3 15 57.7 & -55 06 02& 1.92& --&  --&       -- &     --  &--&--& --&--& --&--&                    \\ 
300A& 3 15 01.8 & -55 14 08& 4.93& 10.21&  12.96&  S&0.000 &111&--& --&--& 0.00$_{-0.00}^{+0.01}$&--&                   \\ 
308A& 3 13 17.8 & -55 15 52& 1.71& --&  22.13&       -- &    --  &  --  &--& --&--&--&--&                    \\ 
310A& 3 14 57.9 & -55 13 23& 3.21& 18.51&  22.41&  B&1.01 &111   &43.07&-20.72& 1.78&0.05$_{-0.05}^{+0.20}$ & 43.09&                   \\ 
311A& 3 16 29.8 & -55 11 02& 3.88& 18.61&  --&       -- &     --  & --  &--& --&--&--&--&                    \\ 
334A& 3 16 31.6 & -55 12 27& 0.70& 19.32&  20.17&  B&2.536 &1\,-\,-   &44.73&-25.13& 1.93&0.52$_{-0.52}^{+0.69}$ & 44.79&       2,3\\
335A& 3 13 57.3 & -55 04 32& 0.91& 15.96&  17.97&  N&0.257 &111&41.93& --&--& 1.91$_{-0.96}^{+1.77}$ & 42.54&                   \\ 
345A& 3 13 16.6 & -55 22 38& 0.84& --&  21.61&  B&2.166 &111&44.88&-23.36& 1.54&0.00$_{-0.00}^{+0.10}$ & 44.88&                    \\ 
360A& 3 13 11.7 & -55 17 18& 2.36& 20.02&  --&  B&2.166 &111&44.49& --&--&0.14$_{-0.14}^{+0.90}$ & 44.51&                   \\ 
361A& 3 13 14.6 & -55 03 18& 2.03& --&  20.03&  B&1.648 &111&44.20&-24.28& 1.96&12.0$_{-6.4}^{+9.7}$ & 44.80&                   \\ 
364A& 3 14 12.0 & -55 25 56& 1.82& --&  21.10&  B&1.821 &1\,-\,-&44.39&-23.48& 1.75&0.23$_{-0.23}^{+0.36}$ & 44.45&        2,3\\ 
367A& 3 16 34.8 & -55 03 49& 3.92& --&  21.08&  B&0.981&1\,-\,-&43.97&-21.99& 1.61&2.29$_{-1.46}^{+1.97}$ & 44.37&              2\\ 
371A& 3 16 20.5 & -55 03 33& 3.16& --&  21.25&  N&0.802 &110   &43.23& --&--&33.1$_{-10.6}^{+20.5}$ & 44.37&                   \\ 
373A& 3 15 02.3 & -55 05 07& 2.53& --&  23.69&  B&0.953&101&43.24&-19.33& 1.45&0.00$_{-0.00}^{+0.06}$ & 43.24&                    \\ 
373B& 3 15 02.5 & -55 05 02& 6.11& --&  21.87&        N &      0.573&010  &--& --&--& --&--&                    \\ 
377A& 3 16 53.4 & -55 08 38& 0.81& --&  22.91&       -- &     --  &--&--& --&--&  --&--&                   \\ 
381A& 3 16 44.3 & -55 27 08& 2.93& --&  19.34&  B&1.879 &1\,-\,-&45.01&-25.32& 1.80&0.00$_{-0.00}^{+0.07}$ & 45.01&               1\\ 
382A& 3 12 47.3 & -55 16 51& 1.46& 19.83&  20.63&  B&1.904&111    &44.57&-24.07& 1.78&0.00$_{-0.00}^{+0.04}$ & 44.57&          C         \\ 
388A& 3 14 01.3 & -54 59 56& 0.72& 17.27&  19.99&  B&0.841 &1\,-\,-&44.74&-22.75& 1.39&0.00$_{-0.00}^{+0.01}$ & 44.74&                1\\ 
408A& 3 15 54.6 & -55 10 04& 4.12& 19.92&  23.40&       -- &     --  &--&--& --&--&--&--&                     \\ 
410A& 3 15 02.4 & -55 27 45& 4.21& 11.01&  14.68&  S&0.000 &111&--& --&--& 0.00$_{-0.00}^{+0.01}$ &--&                   \\ 
422A& 3 15 31.6 & -55 10 45& 4.79& 17.22&  21.04&  N&0.476 &111&42.50& --&--&3.98$_{-1.61}^{+2.46}$ & 43.21&         \\ 
437A& 3 15 33.7 & -55 02 58& 2.33& 20.22&  21.78&  S&0.00 &111&--& --&--&0.01$_{-0.01}^{+0.05}$&--&                    \\ 
449A& 3 15 43.8 & -55 07 42& 4.28& 18.62&  --&  B&1.204 &111&43.47& --&--&7.59$_{-4.04}^{+6.10}$ & 44.08&                3\\ 
462A& 3 15 06.0 & -55 16 24& 1.99& 18.75&  --&       -- &     --  &--& --&   --&--& --&--&                     \\ 
463A& 3 16 25.3 & -55 08 39& 0.55& 19.07&  --&  N&2.531&111    &44.41& --&--&3.63$_{-2.20}^{+4.27}$ & 44.66&                  \\ 
473A& 3 15 32.2 & -55 11 22& 6.40& --&  21.64&        S &      0.000  &011&--& --&--&0.01$_{-0.01}^{+0.05}$&--&                   \\ 
473B& 3 15 32.9 & -55 11 27&11.67& --&  17.22&        G &      0.154  &0\,-\,-&--& --&--&--&--&               7\\ 
475A& 3 13 52.4 & -55 03 46& 1.77& 20.20&  23.97&  B&1.729 &111&44.07&-20.48& 1.36& 0.10$_{-0.10}^{+0.43}$ & 44.10&                   \\ 
480A& 3 16 53.2 & -55 12 32& 2.52& --&  22.47&  N&0.921&100&43.52& --&--& 4.79$_{-2.34}^{+3.85}$ & 44.12&                    \\ 
485A& 3 15 27.1 & -55 16 14& 4.69& 17.75&  22.85&  N&1.263&100&43.33& --&--&27.5$_{-15.1}^{+72.5}$ & 44.25&                    \\ 
496A& 3 12 45.4 & -55 16 48& 1.38& --&  19.74&  B&1.906&111    &44.63&-24.96& 1.91&0.00$_{-0.00}^{+0.14}$ & 44.63&         C          \\ 
498A& 3 15 06.1 & -55 14 00& 1.09& 11.83&  13.65&  S&0.00 &111   &--& --&--&0.01$_{-0.01}^{+0.06}$ & -- &                    \\ 
511A& 3 15 28.0 & -55 13 14& 0.92& 19.88&  22.58&  B&2.623 &111&44.22&-22.79& 1.75&0.00$_{-0.00}^{+0.30}$ & 44.22&           \\ 
512A& 3 13 48.2 & -55 13 09& 1.12& 17.90&  --&  N&0.465 &110&42.83& --&--&1.91$_{-1.02}^{+1.53}$ & 43.36&             C\\ 
512B& 3 13 48.3 & -55 13 05& 3.57& 15.92&  --&        N &      0.584&0\,-\,-    &--& --&--& --&--&              3,C\\ 
518A& 3 14 57.0 & -55 29 25& 2.01& --&  21.53&       -- &     --  &--&--&    --&--& --&--&                   \\ 
524A& 3 13 14.1 & -55 04 16& 3.54& --&  20.46&        S &      0.000  &011&--& --&--&--&--&  \\ 
531A& 3 13 38.2 & -55 03 51& 1.12& 18.31&  --&  B&0.926 &111&44.15& --&--&0.00$_{-0.00}^{+0.01}$ & 44.15&                    \\ 
535A& 3 16 53.5 & -55 11 52& 1.51& --&  21.81&  B&0.974&101&43.43&-21.25& 1.71&0.44$_{-0.31}^{+0.58}$ & 43.60&                    \\ 
536A& 3 14 50.7 & -55 04 10& 3.47& --&  22.41&  N&0.333 &111&41.92& --&--&--&--&                   \\ 
543A& 3 12 50.3 & -55 11 00& 0.56& 19.85&  --&  B&2.511&111    &44.44& --&--&--&--&                    \\ 
551A& 3 12 43.3 & -55 12 01& 3.16& --&  --&        S &      0.000 &011 &--& --&--& --&--&                     \\ 
551B& 3 12 43.0 & -55 11 58& 1.20& --&  --&  B&0.789 &111&43.48& --&--&0.00$_{-0.00}^{+0.20}$ & 43.48&                   \\ 
557A& 3 14 41.4 & -55 21 37& 1.27& 15.42&  18.07&  G&0.16 &111&41.26& --&--&0.00$_{-0.00}^{+0.02}$ & 41.26&                    \\ 
559A& 3 16 17.3 & -55 14 29& 0.99& 17.95&  21.95&  B&1.180 &1\,-\,-&43.82&-21.51& 1.60&0.00$_{-0.00}^{+0.02}$ & 43.82&                3\\ 
579A& 3 15 47.3 & -55 14 19& 0.88& 18.01&  20.84&  N&0.497&110    &42.43&--& --&0.23$_{-0.12}^{+0.18}$ & 42.60&                   \\ 
582A& 3 16 30.5 & -55 11 30& 0.89& 17.23&  21.32&  S&0.000 &111& --&--&--&0.04$_{-0.04}^{+0.06}$&                     \\ 
585A& 3 13 09.3 & -55 11 40& 1.80& 19.23&  23.82&  B&1.383 &111&44.29&-20.01& 1.16&0.01$_{-0.01}^{+0.20}$ & 44.30&                     \\ 
593A& 3 15 11.2 & -55 12 02& 1.16& 18.61&  22.12&  B&0.964 &111&43.19&-20.92& 1.75&0.00$_{-0.00}^{+0.13}$ & 43.19&                     \\ 
607A& 3 16 51.2 & -55 13 06& 0.22& --&  19.83&  N&0.407 &110&42.87&--&--&0.00$_{-0.00}^{+0.01}$ & 42.87&      C               \\ 
608A& 3 13 49.7 & -55 12 59& 0.57& 17.67&  22.46&  N&1.019&100&43.58& --&--& 0.14$_{-0.14}^{+0.24}$ & 43.64&              4\\ 
610A& 3 15 51.8 & -55 12 22& 1.25& 17.64&  21.12&  N&0.699 &111&43.10& --&--& 8.32$_{-2.45}^{+3.14}$ & 43.90&                   \\ 
615A& 3 14 59.5 & -55 03 57& 1.75& 19.15&  22.77&  B&1.355&101&43.99&-21.01& 1.46& 0.00$_{-0.00}^{+0.06}$ & 43.99&     C              \\ 
615B& 3 15 00.2 & -55 03 53& 5.00& 17.08&  19.23&        N &      0.096&011    &--& --&--& --&--&               3, C\\ 
631A& 3 15 32.7 & -55 12 00& 0.89& 19.65&  --&       -- &     --  & --  &--& --&--&  --&--&                   \\ 
632A& 3 16 24.6 & -55 21 33& 2.07& 18.97&  23.29&        ? &    --  &000&--& --&--&  --&--&       C             \\ 
632B& 3 16 24.7 & -55 21 44& 9.08& 19.04&  22.93&        N &      0.987  &011&--& --&--& --&--&                    \\ 
652A& 3 15 30.3 & -55 04 37& 0.52& 18.56&  20.75&  B&2.710&111 &45.14&-24.66& 1.68&0.00$_{-0.00}^{+0.20}$ & 45.14&               3\\ 
653A& 3 16 47.4 & -55 12 31& 1.38& --&  23.38&       -- & -- &     --  &--& --&--& --&--&               4,C\\ 
654A& 3 17 05.7 & -55 27 18& 3.54& --&  --&  B&2.105 &1\,-\,-&45.26& --&--&0.33$_{-0.33}^{+0.44}$ & 45.32&              1\\ 
656A& 3 13 26.1 & -55 04 25& 1.36& 15.35&  18.37&  N&0.166 &111&42.64& --&--&0.91$_{-0.32}^{+0.34}$ & 43.13&                   \\ 
664A& 3 16 10.0 & -55 21 20& 9.50& 18.31&  22.60&       -- &-- &     --  &--& --&--&  --&--&                   \\ 
664B& 3 16 11.0 & -55 21 22& 5.17& --&  --&       -- &     --  &--  &--& --&--&   --&--&                  \\ 
673A& 3 15 40.3 & -55 12 21& 0.65& 18.76&  22.00&  B&1.062 &111&44.03&-21.21& 1.45&0.40$_{-0.19}^{+0.29}$ & 44.18&             3\\ 
675A& 3 16 24.6 & -55 11 44& 0.66& 19.14&  21.82&  B&1.140 &1\,-\,-&44.06&-21.55& 1.50&0.00$_{-0.00}^{+0.06}$ & 44.06&               3\\ 
686A& 3 13 46.6 & -55 11 49& 0.67& 17.08&  19.11&  B&1.663 &111&45.44&-25.23& 1.58&0.00$_{-0.00}^{+0.03}$ & 45.44&  1,3,5\\ 
687A& 3 13 02.6 & -55 26 08& 4.54& --&  18.83&        S &      0.000&011   &--& --&--& --&--& \\
\end{longtable}
\end{landscape}


\begin{thebibliography}{} 
 


 

  \bibitem[Alexander et al.(2003)]{alexander} Alexander, D.M., Bauer, F.E.,  
      Brandt, W.N., et al. 2003, \aj, 125, 383 
  

  \bibitem[Akylas et al.(2006)]{akylas}  Akylas, A., Georgantopulos, I.,
           Georgakakis, A., Kitsionas, S., and Hatziminaoglou, E. 2006, \aap, in press
      
  \bibitem[Avni \& Tananbaum(1986)]{avni} Avni, Y., \& Tananbaum, H., 1986, ApJ, 305, 83

 
   \bibitem[Brandt \& Hasinger(2005)]{brandt} Brandt, W.N., \& Hasinger, 
     G. 2005, \araa, 43, 827 
 
 
  \bibitem[Caccianiga et al.(2004)]{caccianiga} Caccianiga, A., 
       Severgnini, P., Braito, V., et al. 2004, \aap, 416, 901  


 
  \bibitem[Cash(1979)]{cash} Cash, W. 1979, \aj, 228, 939


  \bibitem[Ciliegi et al.(2003)]{ciliegi}  
       Ciliegi, P., Zamorani, G., Hasinger, G., et al. 2003, \aap, 398, 901        

  \bibitem[Comastri et al.(2002)]{comastri} Comastri, A., Mignoli, M., 
        Ciliegi, P., et al. 2002, \aj, 571, 771  

 \bibitem[Della Ceca et al.(2004)]{dellaceca} Della Ceca, R., Maccacaro, T.,
        Caccianiga, A., et al. 2004, \aap, 428, 383


  \bibitem[Dwelly \& Page(2006)]{dwelly} Dwelly, T., \& Page, M.J.  
       2006, \mnras, 372, 1755 


  \bibitem[Francis et al.(1991)]{francis} Francis, P.J., Hewett, P.C.,  
       Foltz, C.B., et al. 1991, \apj, 373, 465 



  \bibitem[Green et al.(1995)]{green0} Green, P.J., Schartel, N., Anderson, S.F., et al. 1995, ApJ, 450, 51

  \bibitem[Green et al.(2004)]{green} Green, P.J., Silverman, J.D.,  
      Cameron, R.A., et al. 2004, \apjs, 150, 43 

   


  \bibitem[Gruppioni et al.(1999)]{gruppioni} Gruppioni, C., Mignoli, M.,  
      Zamorani, G. 1999, \mnras, 304, 199 

 
  \bibitem[Gruppioni et al.(1997)]{gruppioni1} Gruppioni, C., Zamorani, G.,  
      de Ruiter, H.R., et al. 1997, \mnras, 286, 470 

 
  \bibitem[Hasinger et al.(2001)]{hasinger1} Hasinger, G.,  
      Altieri, B., Arnaud, M., et al. 2001, \aap, 365, 45 

 
 
  \bibitem[Hasinger et al.(1998)]{hasinger} Hasinger, G.,  
       Burg, R., Giacconi, R., et al. 1998, \aap, 329, 482 

 
   \bibitem[Horne(1986)]{horne} Horne, K. 1986, PASP, 98, 609 


  \bibitem[Isobe et al.(1990)]{isobe} Isobe, T., 
      Feigelson, E.D., Akritas, M.G. 1990, ApJ, 364, 104


  \bibitem[Kennicutt(1992)]{kennicutt} Kennicutt, C.K. 1992, ApJ, 388, 310


  \bibitem[Komossa et al.(1998)]{komossa} Komossa, S., 
      Schulz, H., Greiner, J. 1998, \aap, 334, 110


  \bibitem[La Franca et al.(2002)]{lafranca} La Franca, F., Fiore, F.,  
       Vignali, C., et al. 2002, \apj, 570, 100  


 \bibitem[Lamer et al.(1997)]{lamer} Lamer, G., Brunner, H.,
      Staubert, R. 1997, \aap, 327, 467
 
 \bibitem[Lampton et al.(1976)]{lampton} Lampton, M., Margon, B.,  
      Bowyer, S. 1976, \aj, 208, 177 

 
  \bibitem[Lehmann et al.(2001)]{lehmann} Lehmann, I., Hasinger, G.,  
      Schmidt, M., et al. 2001, \aap, 371, 833 

 
 
 \bibitem[Mainieri et al.(2002)]{mainieri} Mainieri, V., Bergeron, J., 
      Hasinger, G., et al. 2002, \aap, 393, 425 


 
  \bibitem[Marano et al.(1988)]{marano} Marano, B., Zamorani, G.,  
      Zitelli, V. 1988, \mnras, 232, 111 


  \bibitem[Mateos et al.(2005)]{mateos} Mateos, S., Barcons, X., Carrera,
  F.J., et al. 2005, \aap, 444, 79 

 

  \bibitem[Mignoli \& Zamorani(1998)]{mignoli} Mignoli, M., \& Zamorani, G., 
      1998, ``The Young Universe'', ASP Conference Series, 146, 80 



  \bibitem[Oke \& Gunn(1983)]{oke} Oke, J.B., \& Gunn, J.E. 1983,
      \aj, 266, 713 


    \bibitem[Osborne(2001)]{osborne} Osborne J. 2001, SSC-LUX-TN-0059 (issue3),  
      http:\,-\,-xmmssc-www.star.le.ac.uk/pubdocs/SSC-LUX-TN-0059\_3.ps.gz 



  \bibitem[Rosati et al.(1998)]{rosati} 
    Rosati, P., Della Ceca, R., Norman, C., and
    Giacconi, 1998, \apj, 492, L21


  \bibitem[Sarazin(1997)]{sarazin} Sarazin, C.L. 1997,  
      ASP Conference Series, 116, 375 


  \bibitem[Severgnini et al.(2003)]{severgnini} Severgnini, P., 
      Caccianiga, A., Braito, V., et al. 2003,  
      \aap, 406, 483 


  \bibitem[Silverman et al.(2005)]{silverman} Silverman, J.D.,  
      Green, P.J., Barkhouse, W.A., et al. 2005,  
      \apj, 618, 123 


  \bibitem[Steffen et al.(2006)]{steffen} Steffen, A.T., Strateva, I., Brandt, W.N., et al. 2006, 
          \apj, 131, 2826
 
  \bibitem[Sutherland \& Saunders(1992)]{sutherland} Sutherland, W., \&  
      Saunders, W. 1992, \mnras, 259, 413


  \bibitem[Szokoly et al.(2004)]{szokoly} Szokoly, G.P.,  
      Bergeron, J., Hasinger, G., et al. 2004, 
      \apjs, 155, 271 

  
  \bibitem[Teplitz et al.(2003)]{teplitz} Teplitz, H.I., 
      Collins, N.R., Gardner, J.P., et al. 2003 
      \apjs, 146, 209 

 
 
  \bibitem[Tozzi et al.(2003)]{tozzi} Tozzi, P., 
      Gilli, R., Mainieri, V., et al. 2006,  
      \aap, 451, 457 



  \bibitem[Vanden Berk et al.(2001)]{vandenberk} Vanden Berk, D.E.,  
      Richards, G.T., Bauer, A., et al. 2001, 
      \aj, 122, 549 

 
  \bibitem[White \& Davis(1998)]{white} White III, R.E., \& Davis, S.D. 1998 
      ASP Conference Series, 136, 299 

 
  \bibitem[Worsley et al.(2005)]{worsley} Worsley, M.A., Fabian, A.C., Bauer, 
      F.E., et al. 2005, \mnras, 357, 1281 



  
  \bibitem[Zamorani et al.(1999)]{zamorani} Zamorani, G., Mignoli, M. 
      Hasinger, G., et al. 1999,  
      \aap, 346, 731 

 
  \bibitem[Zitelli et al.(1992)]{zitelli} Zitelli, V., Mignoli, M. 
      Zamorani, G., et al. 1992,  
      \mnras, 256, 349 

 
  \bibitem[Zombeck(1990)]{zombeck} Zombeck, M.V. 1990,  
      Handbook of Astronomy, Cambridge University Press
     


 

\end{thebibliography}
\end{document}